\documentclass[10pt, format=manuscript,nonacm]{acmart}

\AtBeginDocument{%
  }

\usepackage{geometry}
\usepackage{multirow}
\usepackage{graphicx}
\usepackage{subcaption}
\usepackage[linesnumbered,ruled,vlined,noend]{algorithm2e}
\usepackage{cleveref}

\begin{document}
	
	\title{Multi-Scenario User Profile Construction via Recommendation Lists}
	
	\author{Hui Zhang}
	\email{D202481208@hust.edu.cn}
		\affiliation{%
		\institution{Huazhong University of Science and Technology}
		\city{Wuhan}
		\state{Hubei}
		\country{China}
	}
    \author{Jiayu Liu}
	\email{jiayuliu@hust.edu.cn}
		\affiliation{%
		\institution{Huazhong University of Science and Technology}
		\city{Wuhan}
		\state{Hubei}
		\country{China}
	}

	\begin{abstract}
	Recommender systems (RS) play a core role in various domains, including business analytics, helping users and companies make appropriate decisions. To optimize service quality, related technologies focus on constructing user profiles by analyzing users' historical behavior information. This paper considers four analytical scenarios to evaluate user profiling capabilities under different information conditions. A generic user attribute analysis framework named RAPI is proposed, which infers users' personal characteristics by exploiting easily accessible recommendation lists. Specifically, a surrogate recommendation model is established to simulate the original model, leveraging content embedding from a pre-trained BERT model to obtain item embeddings. A sample augmentation module generates extended recommendation lists by considering similarity between model outputs and item embeddings. Finally, an adaptive weight classification model assigns dynamic weights to facilitate user characteristic inference. Experiments on four collections show that RAPI achieves inference accuracy of 0.764 and 0.6477, respectively.

	\end{abstract}
	
	\keywords{Recommender System}

	\maketitle
	
	\section{Introduction}
	
	Recommender systems provide personalized recommendation \cite{dblp1,dblp2,dblp3,dblp4,dblp5} services to alleviate the problem of information overload. Typically, recommender systems analyze the user's historical interaction behaviors (ratings, clicks, views, etc.) to learn a user's preferences and generate an appropriate recommendation list for the user.  However, a more accurate recommendation list is more likely to contain the information that reflects the user's real identity. An business analyst is able to utilize this information to infer the user's history behaviors and personal characteristics (age, gender, occupation, etc.)
	\par
	
	Recently, research has explored the inferential capabilities of recommendation models for user characteristics. Such profiling techniques leverage relational information and interaction history to establish correlations between observed behaviors and specific user attributes. Kosinski et al. ~\cite{kosinski2013private} exploit users’ behavioral information about Facebook likes to infer their private attribute information. However, aforementioned methods primarily focus on system's internal architecture, or query-level access. There remains a research gap in evaluating the associative transparency between system-generated recommendation sequences and specific user characteristics. Recommendation lists can be obtained by various methods: sharing on social media, comments on forums or communities, recommendations from bloggers and websites, etc. Kosinski et al. ~\cite{kosinski2013private} argue that recommendation lists can also be regarded as a kind of digital record.
    \par
	  
	To comprehensively analyze the information embedded in recommendation lists and evaluate the potential for constructing user profiles, this research investigates the subject through several defined roles. In this work, the party constructing user profiles is referred to as the analyst; a user providing interaction history is termed a reference user; and the individual whose profile is being constructed is the target user. We establish four scenarios and utilize profiling accuracy as a quantitative indicator to assess the effectiveness of user profile analysis across different conditions. In all scenarios, the analyst is assumed to have access to the target user's recommendation lists. Importantly, this study focuses on quantifying the extent of user information extractable from these lists, rather than the methods by which the analyst obtains historical engagement records—whether through explicit user consent or authorized third-party channels.
	\par
	
    This study encompasses four distinct analytical scenarios representing varying levels of information accessibility. Scenario 1 (External Observer) characterizes the analyst as an external entity with access limited to publicly observable recommendation outputs. Scenario 2 (Platform Insider) positions the analyst as an internal developer with privileged access to system architectures and latent embedding representations. Scenario 3 (Third-party Analyst) involves an external researcher employing content-based inference methodologies with constrained system accessibility. Scenario 4 (Comprehensive Analyst) represents an advanced analyst capable of aggregating interaction records from reference users through multiple authorized information channels. These scenarios collectively enable systematic evaluation of the effectiveness of user profile construction under varying information availability conditions.In Scenarios 1 and 2, the demographic attributes of target users can be obtained by classifying the item numbers or item embedding by an advanced classification model, respectively. However, Scenarios 3 and 4 face some challenges: (1) the analyst will not be satisfied with the original recommendation model that is blackbox for him, which would greatly limit the analyst's ability; (2) the analyst needs more recommendation information from the user for higher accuracy, while user's more recommendation information is hard to obtain. To solve the first challenge, an alternative model which is whitebox for the analyst is established based on the recommendation list similarity. To overcome the second challenge, an information augmentation module is introduced to generate supplementary recommendation outputs. This is achieved by calculating embedding similarities and integrating content embeddings derived from a pre-trained BERT model \cite{bert}.
	\par
	
    In this work, a generalized User Profile Analysis framework called RAPI (Recommendation Analysis for Profile Inference) is proposed, which can utilize a user's recommendation list along with optional interaction records to accurately characterize user attributes. The RAPI framework is divided into mainly three steps. Firstly, the analyst constructs a surrogate recommendation model, which not only achieves recommendation performance similar to the original system but also operates as a whitebox for the analyst, implying that the analyst can efficiently access the embedding representations of the reference user and their interacted items. Secondly, the analyst designs an information augmentation module. It computes the similarity between a user's original recommendation list and remaining items of the item pool, subsequently adds a subset of the most similar items to the original recommendation list. As the surrogate recommendation model has no access to item embeddings with which reference users have not interacted, a pre-trained language model, BERT, is employed to generate content embeddings for these items. The content embeddings are then aligned to recommendation embeddings in the same latent space by an alignment model. Finally, an adaptive-weight classification model is designed to accurately infer the user's attributes based on the augmented recommendation list.
	\par
	This framework is evaluated on four real recommendation outputs. The experimental results indicate that the RAPI can characterize user attributes with higher accuracy compared with baseline methods, even when prior knowledge is limited to partial interaction records or entirely absent. This study underscores the potential for user attributes to be inferred from recommendation lists or interaction records, highlighting the importance of understanding user characteristic modeling in recommendation systems. Overall, our contributions in this paper are as follows:
	\begin{itemize}
		\item Four scenarios are designed to demonstrate the effectiveness of user profile analysis. To the best of our knowledge, this work is the first to systematically examine the inference of user attributes from recommendation lists.
		\item RAPI is proposed to infer user attributes from recommendation lists. Specifically, RAPI constructs a surrogate model based on the similarity between recommendation lists, performs recommendation list augmentation based on the similarity between recommendation embeddings, and dynamically assigns weights to items at different positions to enhance accuracy of inferences.
		\item  Extensive experiments on four outputs demonstrate the effectiveness of RAPI, which can characterize user attributes with high accuracy.
	\end{itemize}

    \section{Related Work}
	\subsection{Collaborative Filtering}
	Collaborative filtering tasks are commonly approached through the application of matrix factorization methodologies in recommendation systems \cite{mf, mf1, mf2, mf3, dl1, dl2, dl3, gcn1, gcn2, gcn3, lightgcn, neumf, ngcf}. The core concept behind this approach is to project input signals, e.g. user and item numbers, into a unified latent space where their similarities are quantified by an interaction function. In its most elementary form, this function is represented by a simple dot product, effectively capturing the degree of interaction between the embedded user and item vectors \cite{mf}. Given the remarkable success of this foundational method, numerous adaptations of matrix factorization have since been devised, each seeking to refine and enhance the original method's capabilities \cite{mf1, mf2, mf3}. Probabilistic matrix factorization \cite{mf2} is a bayesian approach to matrix factorization that models user-item interactions as a product of low-rank matrices with Gaussian priors, effectively handling sparsity and cold-start problems in recommendation systems.
	
	Recently, the emergence of deep learning has catalyzed a rapid evolution in collaborative filtering methodologies. Owing to its formidable modeling prowess, deep learning has been extensively integrated into recommendation systems. It serves to cultivate more effective user and item embedding, as well as to capture the intricate dynamics of user-item interactions with enhanced precision and complexity \cite{dl1, dl2, dl3}. For instance, the NeuMF model utilizes deep neural networks to capture the sophisticated interactions between users and items \cite{neumf}. Similarly, A3NCF \cite{dl1} leverages deep learning to discern user attention from reviews, thereby enhancing the recommendation process through a more attentive and personalized approach.
	
	More recently, the integration of Graph Convolution Network (GCN) techniques into recommendation systems has garnered significant attention and has yielded impressive results \cite{gcn1, gcn2, gcn3, ngcf, lightgcn}. The success of GCN in recommendation systems is primarily attributed to its ability to leverage high-order proximity within the user-item interaction graph, which enhances the learning of user and item embedding. IMP-GCN model \cite{gcn1} develops user and item embedding through the execution of high-order graph convolution within sub-graphs that are crafted based on the interests of the users. 
	LightGCN \cite{lightgcn} simplifies GCN for recommendations by removing non-essential operations, offering a simpler, more effective model with significant performance improvements.

    \subsection{User Profile Analysis in Recommender Systems} Recently, a multitude of user profile analysis methods have emerged, contributing to the understanding and modeling of users in recommendation systems, e.g. member behavior analysis, attribute inference methods. The purpose of \textbf{attribute inference methods} in recommender systems is to characterize users' information \cite{2}, e.g. user attributes \cite{3,4,5,6,7,8,10}. Attribute inference methods can be divided into three main categories: one category utilizes information about the target user's friends \cite{3,4,5} and community member \cite{6} to infer the target user's attributes. He et al. \cite{4} employs a bayesian network to analyze the factors affecting attribute inference accuracy and demonstrate that user attributes can be effectively characterized, particularly in social structures. The second group of methods uses behavioral information of users, such as movie rating behaviors \cite{7} to infer information about their attributes. Weinsberg et al. \cite{7} discuss the ability of recommender systems to characterize user demographics like gender from ratings alone, and introduce techniques to enhance this characterization without compromising recommendation quality. The third group of methods utilize both behavioral information \cite{2,8,10} and friend's information. Gong et al. \cite{8} builds a social-behavioral-attribute network that integrates all users' behaviors and friendship information in a unified framework to infer user attributes through a voting distribution model. 
    \par 
   
    Recent research has explored recommender systems' capability for \textbf{membership behavior analysis}, which can ascertain whether an individual's record contributes to model understanding, leading to significant insights for user modeling \cite{mem1,memccs,memseqrec}. VLIA \cite{memccs} leverages both embedding and proximity information of location to analyze user behaviors and visited locations, even when the federated recommendation system is protected with stochastic noise injection. Yuan et al. present a systematic investigation of the characteristics of users in federated recommendations, introducing an interaction-level behavior analysis method, and examine the effectiveness  \cite{mem1}.

	\section{Preliminary}
	\subsection{Notation Definition}
	The user set \(U\), item set \(V\) are noted as $U=\left \{ u_{1},\dots ,u_{M}  \right \} $, $V=\left \{ v_{1},\dots ,v_{N}  \right \} $, where M and N are the number of user set and item set, respectively. The sensitive attribute is noted as  $S=\left \{ s_{1},\dots ,s_{M}  \right \} $, \(s_{i}\) can be gender, occupation, age and so on. The users' interaction history records are noted as $H=\left \{ h_{1},\dots ,h_{M}  \right \} $, where $h_{i}\in \mathbb{R}^{n}$ represents which items a user $u_i$ interacts with. The users' recommendation lists generated by a original recommender system are noted as $Z$, where $z_{i} \in \mathbb{R}^{K}$ represents top-K recommendation list of \(u_{i}\). $\mathcal{F}_{rec}$, $\mathcal{F}_{spu}$, $\mathcal{F}_{ali}$, $\mathcal{F}_{cla}$ stand for original recommendation model, spurious recommendation model, latent embedding alignment model and classification model, respectively. Item set \(V\) has three latent embedding matrices, one for recommendation embedding obtained from $\mathcal{F}_{rec}$ which is noted as $\mathbf{E^R}  \in \mathbb{R} ^{N*L_{1} } $ , one for content embedding which is noted as \(\mathbf{E^C} \in \mathbb{R}^{N*L_{2}}\), and one for recommendation embedding obtained from $\mathcal{F}_{spu}$ which is noted as $\mathbf{E^S}  \in \mathbb{R} ^{N*L_{3} } $ , where \(L_{1}\) , \(L_{2}\) and \(L_{3}\) represent embedding size, respectively.
	\subsection{Scenario Introduction}
	\begin{figure}[t]
		\centering
		\includegraphics[width=\linewidth]{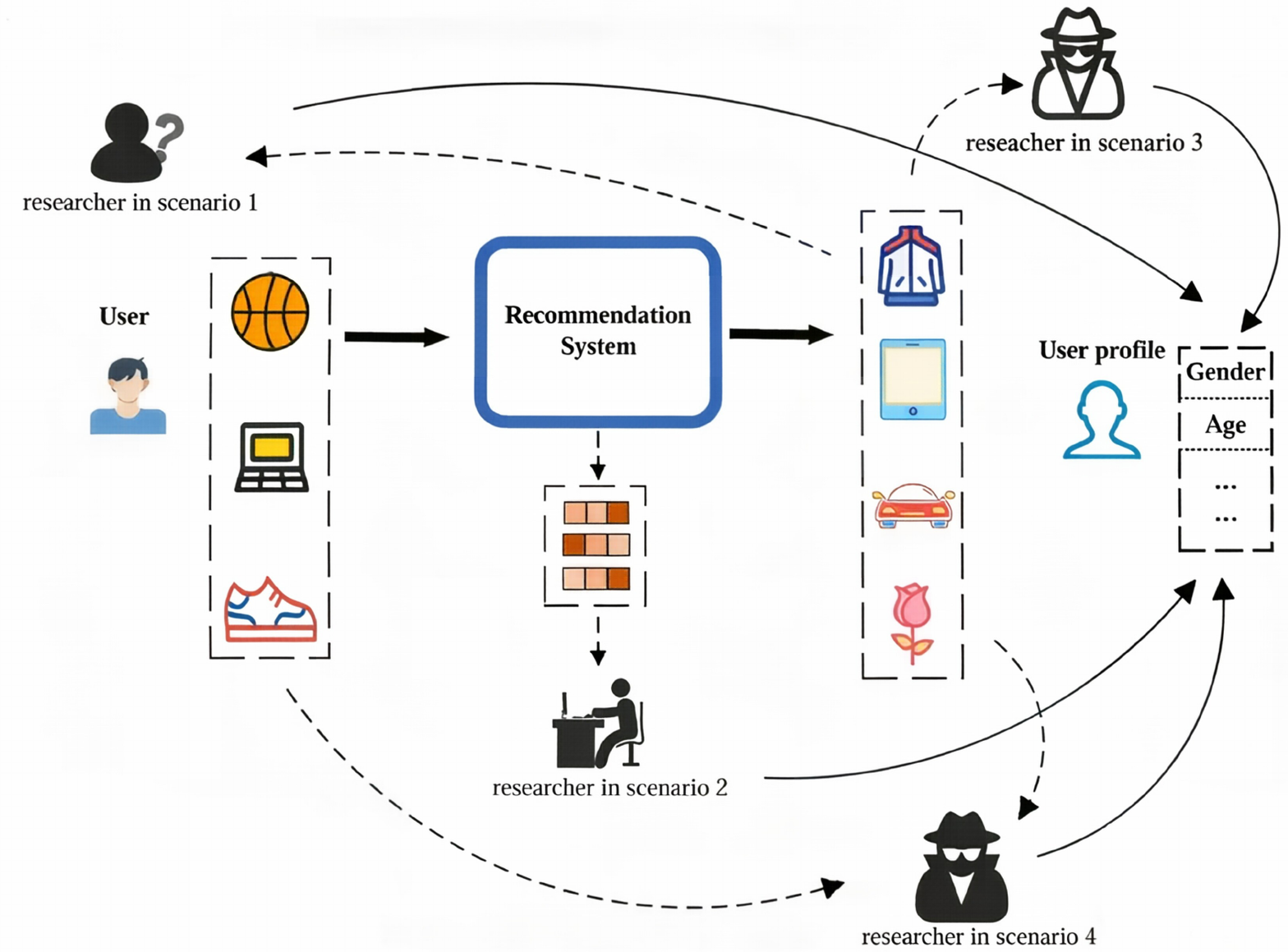}
		\caption{The illustration of four scenarios}
		\label{Fig: scenario}
	\end{figure}

	In our setting, a user who provides interaction history is referred as a "contributor", a user whose attributes are characterized by \textit{RAPI} is referred as a "target user", and the person who performs attribute analysis is referred as an "analyst".The user set $U$ can be divided into contributor set $U^L$ and target user set $U^V$. The ultimate task of the analyst is to characterize the attributes of the target user, however, the analyst's ability to do so varies with the amount of accessible information. Therefore, four different analysis scenarios are set up which are illustrated in Figure \ref{Fig: scenario}, presenting quantity of accessible information for analyst with different identities: 
	\begin{enumerate}
		\item[(1)] the analyst as a curious recommender system user.
		\item[(2)] the analyst as a recommender system developer.
		\item[(3)] the analyst as a researcher with the ability to gather item information. 
		\item[(4)] the analyst as an advanced researcher with additional ability to gather engagement history.
	\end{enumerate} 

	\subsubsection{Scenario 1}
	The analyst is a curious user of the recommender system who wants to characterize attributes from the target user's recommendation lists. That is, all the analyst knows are the target user's recommendation lists and attributes of contributors. This is a reasonable assumption because existing recommender systems often provide recommendation lists as a standard feature, and a user's recommendation list is easily accessible through different methods. Given a target user's recommendation list $z_{i}$, the analyst aims to characterize attributes in the following way:
	\begin{equation}
		s_{i}=\mathcal{F}_{cla}\left (z_{i}   \right )
	\end{equation}
	
	\subsubsection{Scenario 2}
	The analyst is a developer of a recommender system who aims to characterize the target user's attributes through recommendation embedding of items. In contrast to Scenario 1, the analyst also has the ability to leverage the recommendation embedding of items. This represents a stronger assumption, as obtaining the recommendation embedding of items is typically challenging in real-life scenarios. Scenario 2 can be formulated as follows:
	\begin{equation}
		s_{i}=\mathcal{F}_{cla}\left (   AGG \left ( z_{i}   ,\mathbf{E^R } \right )   \right )
	\end{equation}
	where AGG function serves as an aggregation mechanism that amalgamates the embedding associated with items on a recommendation list in various ways. This aggregation process is pivotal in enhancing the performance of attribute analysis by synthesizing the characteristics of all items on the list into a unified representation. More details of AGG are presented in Section \ref{Sec:method}.
	
	\subsubsection{Scenario 3}
	The analyst is a researcher with the ability to gather item information who aims to characterize the target user's attributes through the content embedding of the target user's recommended items. This is a moderate assumption, considering that obtaining the item recommendation embedding is extremely difficult. However, the analyst can obtain the content embedding of the items through item title, text description, etc. Note that any embedding-based approach to representing items is feasible; the approach taken here involves representing items by content embedding derived from BERT. Scenario 3 can be formulated as follows:
	\begin{equation}
		s_{i}=\mathcal{F}_{cla}\left (   AGG \left ( \overline{z_{i}}    ,\mathbf{E^C } \right )   \right )
	\end{equation}
	where \(\overline{z_{i}}\) is the augmented recommendation list, consisting of  \({z_{i}}\) and additional recommendation list generated by spurious model $\mathcal{F}_{spu}$. More details of \(\overline{z_{i}}\) are presented in Section \ref{RLDA} .
	
	\subsubsection{Scenario 4}
	The analyst is a researcher who aims to characterize the target user's attributes using the recommendation embedding of the target user's recommended items. The analyst has the ability to gather contributors' interaction records and recommendation lists. This is a moderate assumption, given the difficulty the analyst faces in obtaining and utilizing the interaction records of contributors. Scenario 4 can be formulated in the following ways:
	\begin{equation}
		s_{i}=\mathcal{F}_{cla}\left (   AGG \left ( \overline{z_{i}}   , \mathbf{E^S } \right )\right )
	\end{equation}
	\par
	Besides the identity information of contributors, it is emphasized that target users' identity information is generally anonymized and never released to the analyst in advance in the above 4 scenarios, even when the analyst is the developer of the recommender system in Scenario 2.

	\section{Method}\label{Sec:method}
	
	In this section, the analyst's analysis methods across various scenarios are outlined. In Scenario 1 and Scenario 2, the analyst can classify the item numbers or the recommendation embedding within the target user's recommendation list to characterize user attributes, respectively. The characterization is facilitated by any advanced classification models, such as Deep Neural Network (DNN) \cite{dnn}. Users' attributes can be characterized by DNN in the following ways:
	\begin{equation}
		\widehat{\mathbf{S} }=softmax\left ( f^{l}\dots f^{2}\left ( f^{1}\left ( \mathbf{A}  \right )  \right )    \right ) 
	\end{equation}
	\begin{equation}
		f^{i}\left (  \mathbf{X}  \right )  =\mathbf{W}^{i}\times \left ( \mathbf{X} \right )^{\top } +\mathbf{B}^{i} 
	\end{equation}
	where \(l\) represents the layers of the DNN, \(f^{i}\) represents the i-layer network of the DNN, \(\mathbf{W}^{i}\) and \(\mathbf{B}^{i} \) represent the weight matrix and bias matrix of the i-th layer, respectively. The embedding matrix $\mathbf{A}$ is items' one-hot embedding matrix in scenarios 1 and recommendation embedding $\mathbf{E}^R$ in scenarios 2.
	\par

\begin{figure*}
		\centering
		\includegraphics[width=\linewidth]{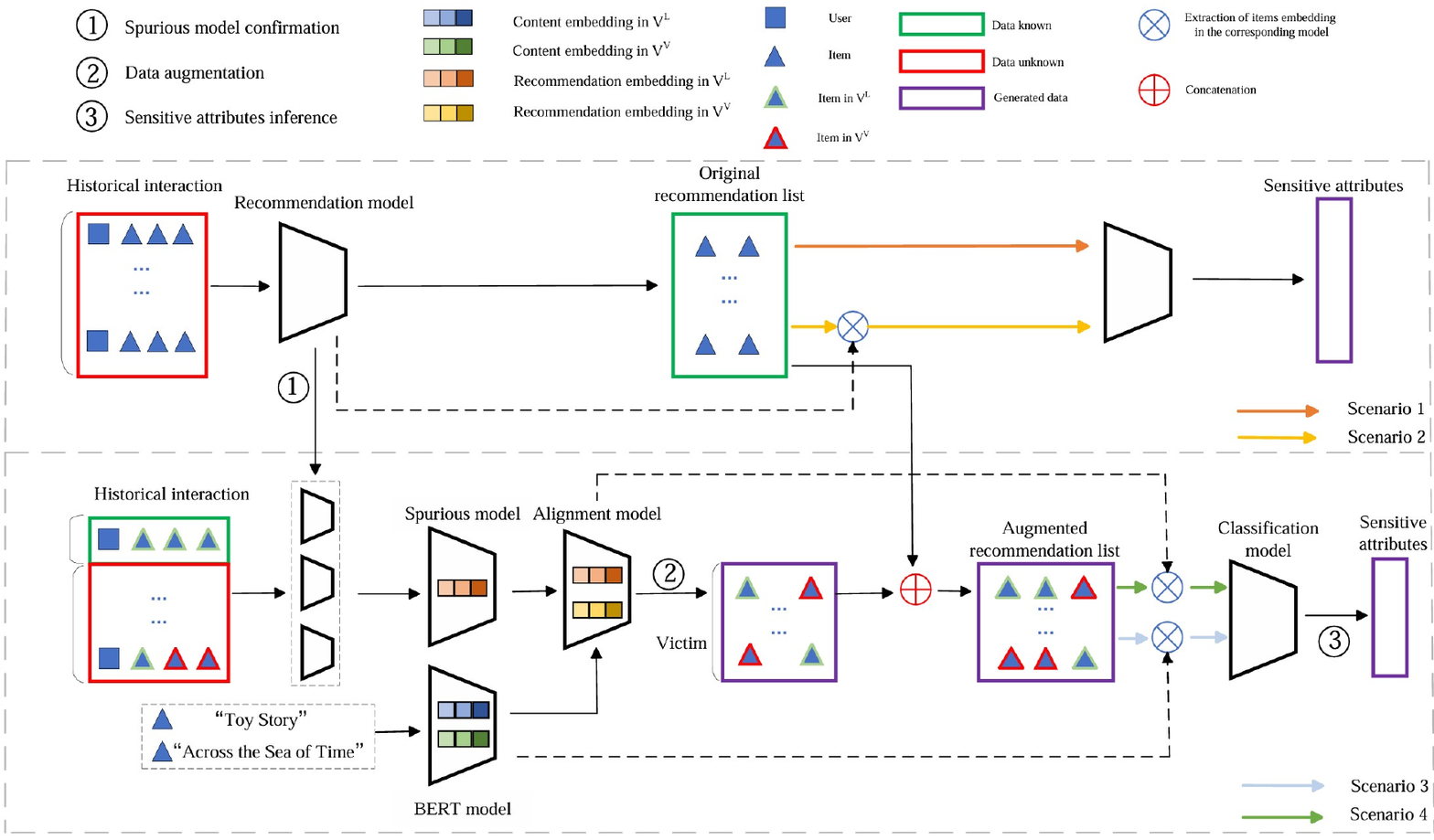}
		\caption{The overall framework of our proposed RAPI. The upper part of the figure illustrates the analyst's actions for characterizing user attributes in Scenario 1 and 2. The lower part of the figure illustrates the analyst's actions for characterizing user attributes in Scenario 3 and 4. }
		\label{Fig: method}
	\end{figure*}

    For Scenarios 3 and 4, a generic framework for user attribute characterization known as \textit{RAPI} (Recommendation Analysis for Profile Inference) is proposed, which primarily consists of three components: a spurious recommendation model confirmation module, a recommendation list augmentation module, and an adaptive weight classification module. The overall framework of \textit{RAPI} is shown in Figure \ref{Fig: method}.
	
	\subsection{ Spurious Model Confirmation  }
	\label{RAPIP1}
	Although the analyst is an external researcher in Scenarios 3 and 4, they have no knowledge of the structure or parameters of the original recommendation model. The employ of model inversion analysis \cite{17} would enable the analyst to have greater operational capability. The analyst can establish a spurious model by leveraging the similarity between spurious recommendation lists and the lists generated by the original recommendation model. In particular, users' interaction records is processed by various recommendation models to produce tailored recommendation lists, which are then utilized to compare with original lists to confirm the spurious model. The process of generating a user's original recommendation list can be formulated as follows:
	\begin{equation}\label{candidate list}
		Z =\mathcal{F}_{rec}(U, V, H)
	\end{equation}
	where $\mathcal{F}_{rec}$ could be any recommender, e.g. LightGCN \cite{lightgcn}. Recommendation list similarity ($\mathit{rls}$) is defined as follows:
	\begin{align}
		\mathit{rls}=&  \sum_{i=1}^{M} \frac{\left | z_{i}\cap \widehat{ z_{i}}  \right | }{\left | z_{i} \right | }, u_i\in U^{L} \label{rls}\\
		\widehat{Z} =& \mathcal{F}_{spu}(U^{L}, V^{L}, H^{L})
	\end{align}
	where \(\widehat{ z_{i}}\) is a recommendation list of user $u_i$ generated by analyst-selected recommendation models, $V^{L}$ is the set of user-related items and $H^L$ is the set of interactions of users. The model with the largest \(\mathit{rls} \) value with the generated recommendation list is confirmed as the final spurious recommendation model.
	
	\subsection{RLDA: Data Augmentation for Recommendation List}\label{RLDA}
	\label{RAPIP2}
	In this section, a \textbf{d}ata \textbf{a}ugmentation method for \textbf{r}ecommendation \textbf{l}ists named \textbf{RLDA} is introduced, which aims at mining more items related to user preference for enhancing the ability of analyst to characterize the attributes of users. Note that the benchmark for RLDA is user's original recommendation lists.
\par
As the surrogate recommendation model is established by analyst, it is a white-box for analyst. The analyst firstly extracts the embedding $\mathbf{E}^S$ of items from the surrogate recommendation model. However, if only the embedding $\mathbf{E}^S$ of the user-related items can be utilized, the augmentation part of the user's recommendation list contains only the user-related items, which obviously affects the accuracy of the inference model. For the items interacted only by users, we proposed to leverage a pre-trained model, BERT, to learn their content embedding $\mathbf{E}^C$ based on their content information, such as title.
\par
However, the two kinds of item embedding $\mathbf{E}^C$ and $\mathbf{E}^S$ are in the different latent spaces. It is hard to calculate the similarity between items crossing two latent spaces. We propose an alignment model to transform the embedding $\mathbf{E}^C$ into the same latent space as $\mathbf{E}^S$. The item set $V$ is divided into two parts: one is user-related items set $V^L$, the other is $V^V$. We use $V^L$ as the training set, $V^V$ as the test set of the alignment model. The alignment model can be formulated as follows:
	\begin{equation}
		f\left ( \mathbf{\widehat{E}}^C, V^L  \right ) = \mathbf{W}\times \left ( f \left ( \mathbf{E}^C,  V^L\right ) \right ) +\mathbf{B}
	\end{equation}
	where \(f \left ( \mathbf{E}^C,  V^L\right )\) implies the operation of extracting the embedding of items in \(V^L\) from the item embedding matrix \(\mathbf{E}^C\) and \(\mathbf{W}\), \(\mathbf{B}\) are the weight and deviation matrices of the alignment model, respectively. \(\mathbf{\widehat{E}}^C\) refers to the content embedding generated by alignment model. The loss function of the alignment model is defined as follows:
	\begin{equation} \label{ali loss}
		\mathcal{L}_T  = -\sum_{i=1}^{M}\sum_{j\in {V}^L} \left ( \widehat{\mathbf{e}} _{j}^{C}-\mathbf{e}_{j}^{C} \right )
	\end{equation}
	After training, the content embedding $\mathbf{E}^C$ of items in $V^V$ is inputed into the alignment model to generate the content embedding $\mathbf{\widehat{E}}^C$. Now all the items' recommendation embedding have stored in the same latent space as follows:
	\begin{equation}
		\mathbf{E}= f \left ( \mathbf{\widehat{E}}^C, V^V \right ) \cup f \left ( \mathbf{E}^{S},  {V}^L\right )
	\end{equation}
	\par
	Since we have obtained the embedding $\mathbf{E}$ of all items, the similarity between the user's historical interactions and candidate items can be measured by Euclidean distance. We define the items that have not been interacted by a user as candidate items. Items with high similarity to the user's historical interactions may better match the user's preference. The similarity \(\mathit{res}\) between a user's original recommendation list $z_i$ and a candidate item $v_b$ is defined as follows:
	\begin{align}
		\mathit{res}(z_i, v_b) &= \frac{1}{K} \sum_{v_a\in z_i} sim(v_{a} , v_{b}) \label{res}\\
		\mathit{sim} \left ( v_{a} , v_{b} \right ) &=\frac{1}{L_{3} }\times \sum_{i=1}^{L_{3}} \left ( \mathbf{e}_{a}\left [ i \right ] - \mathbf{e}_{b}\left [ i \right ]   \right ) \label{res2}
	\end{align}  
	where \(\mathbf{e}_{a}\left [ i \right ]\), \(\mathbf{e}_{b}\left [ i \right ]\) represents the i-th dimension in the embedding \(\mathbf{E}\) of item a, item b, respectively. The candidate items are listed based on the similarity. The top-$K_2$ items $\widehat{z}_i$ are combined with the original recommendation lists as follows:
	\begin{equation}\label{augmented lists}
		\overline{z}_i=concate( z_i,\widehat{z}_i[K_2-K: K_2])
	\end{equation}
	where $\overline{z}_i$ is the augmented recommendation list with length $K_2$.

	\subsection{Adaptive Weight Classification Model}\label{Sec: AGG}
  Based on the augmented recommendation list, items at different positions within the recommendation list may convey varying levels of user preference information. To effectively distinguish between item differences, adaptive weight classification models are utilized to characterize user attributes by assigning different weights to items within the recommendation lists, considering that neither the items in the recommendation list should simply be given the same weight, nor should they be given weights crudely according to their position in the recommendation list, e.g., the higher up the list, the greater the weight. For a user $u_i$, the classification model can be formulated as follows:
\begin{align}
    \widehat{s}_i &= MLP(\mathbf{u}_i) \\
    \mathbf{u}_i &= \sum_{v_j \in z_i} w_j * \mathbf{e}_j^{\mathbf{S}} \label{weight2}
\end{align}
where $\mathbf{e}_j$ is j-th item embedding, $w_j$ is weight for the j-th item within the recommendation list:
\begin{equation} \label{weight1}
    w_{j} = \sum_{k=1}^{K} \frac{exp(W_a * \mathbf{e}_j + b_a)}{exp(W_a * \mathbf{e}_k + b_a)}
\end{equation}
where $\mathbf{W}_a$ and $\mathbf{b}_a$ are trainable parameters, the user attributes $S$ is leveraged as the label to train the classification model. The loss function is Cross Entropy loss which is defined as follows:
\begin{equation} \label{cla loss}
    \mathcal{L}_C = -\frac{1}{|U^L|} \sum_{U^L} \sum^{|S|}_{s} s_i \log \widehat{s}_i
\end{equation}
where $s_i$ and $\widehat{s}_i$ are users' attributes which are regarded as labels and prediction results, respectively.
Algorithm \ref{algorithm} presents an overview of the RAPI framework for user profile construction.

	\begin{algorithm}
		\SetAlgoLined 
		\caption{User Profile Construction for Recommendation Systems } \label{algorithm}
		\KwIn{User set $U$ and item set $V$; original recommendation lists $Z=\left \{ z_{1},\dots ,z_{|\mathit{U}|}  \right \}$; partial interaction history $H $; pre-trained model; candidate spurious models }
		\KwOut{user attributes $S $}
		split $U$ into training user set $U^{L} $ and test user set $U^{V} $ based on whether user interaction history is accessible to the analyst \; 
		split $V$ into interacted item set $V^{L} $ and candidate item set $V^{V} $ based on user interaction records \;
		
		\tcp{ Model Confirmation }
		compute recommendation lists $\mathit{L}$ corresponding to candidate spurious models based on the Equation \ref{candidate list}\;
		
		\For{$i <|\mathit{L}|$}{
			compute $rls_i$ between  $\mathit{l_i}$ and $\mathit{z_i}$ based on the Equation \ref{rls}\;
		}
		$rls=\frac{1}{|L|} {\textstyle \sum_{i=1}^{\left | L \right |}}  rls_i $\;
		confirm spurious model $\mathcal{F}_{spu}$ based on $rls$ and obtain recommendation embedding $\mathbf{E}^S$ in $V^L$\;
		\tcp{Data Augmentation for Recommendation List}
		utilize pre-trained model to extract content embedding $\mathbf{E}^C$ from item title in $V^L$ and $V^V$\;
		train alignment model $\mathcal{F}_{ali}$ with $\mathbf{E}^C$ and $\mathbf{E}^S$ in $V^L$ based on the Equation \ref{ali loss}\;
		utilize $\mathcal{F}_{ali}$ to predict $\mathbf{E}^S$ in $V^V$\;
		compute augmented parts $\widehat{z}$ with $res$ between $z$ and remain items from $\mathbf{E}^S$ based on the Equation \ref{res} and \ref{res2}\;
		compute augmented lists $\overline{z}_i$ by appending $\widehat{z}$ to $z$ based on the Equation \ref{augmented lists}\;
		\tcp{Adaptive Weight Classification Model}
		aggregate attribute embedding by dynamically assigning weights to items at different positions within $\overline{z}_i$ based on the Equation \ref{weight2} and \ref{weight1}\;
		train classification model $ \mathcal{F}_{cla}$ based on the Equation \ref{cla loss}\;
		return  attributes $s$ of users in $U^V$ predicted by	$\mathcal{F}_{cla}$
	\end{algorithm}

	Especially, inspired by the insight that adding perturbance in recommendation lists can improve recommendation diversity, a potential optimization strategy for recommendation systems is proposed, which replaces partial items in recommendation lists with others of the same category. This approach aims to balance recommendation accuracy with content diversity. More details about the optimization strategy are shown in Section \ref{defend}.

	\section{Experiments} \label{experiment}
	In this section, experiments on four real-world recommendation records are conducted to show the effectiveness of proposed RAPI. Experiments mainly focus on the answer to following questions:
	\begin{itemize}
		\item {\textbf{RQ1:}} What is the overall performance of the RAPI framework in inferring sensitive attributes?
		
		\item {\textbf{RQ2:}} What about performance of various modules within the RAPI framework?
		
		\item {\textbf{RQ3:}} How effective is the proposed framework for user profile construction when the original model is a recommendation model considering recommendation quality?
		
		\item {\textbf{RQ4:}} How does the performance of the RAPI framework stack up when utilizing content embedding from a range of pre-trained models?
		
		\item {\textbf{RQ5:}} Does the length of the augmented recommendation list have an effect on the classification model?
		
	\end{itemize}
	
	\subsection{Experimental Settings}
	
	\subsubsection{Datasets}
	Experiments on four publicly available datasets are conducted, \textbf{MovieLens-1m}\cite{mv}, \textbf{LastFM-1k}\cite{LastFM}, Book-Crossing and Cold-Rec. The datasets have been meticulously partitioned into training, testing, and validation subsets with an 8:1:1 ratio, ensuring a robust evaluation framework. The experimental outcomes are derived from the aggregation of three distinct trials, yielding a reliable average that captures the performance consistency of the model. More details of the datasets are shown in table \ref{dataset}.

	\begin{table}
		\caption{Statistical details of the evaluation datasets}
		\label{dataset}
		\begin{tabular}{cccccc}
			\toprule
			Dataset   & \#Users & \#Items & \#Actions & Density & Privacy \\ \midrule
			MovieLens-1m     & 3952    & 6040    & 1000209   & 4.19\%  &
		Gender(2) Age(7)  Occupation(21)  \\
			LastFM-1k & 844     & 71699   & 2248349   & 3.72\% & Gender(2)  \\ 
			Book-Crossing & 275657     & 271179   & 1149746   & 0.002\% & Age(7)  \\
			Cold-Rec & 1546382     & 1391668   & 239834151   & 0.02\% & Gender(2)  \\
				 \bottomrule
		\end{tabular}
	\end{table}
	
	\begin{itemize}
		\item \textbf{MovieLens-1m}  a widely recognized benchmark for collaborative filtering in recommendation systems. This dataset offers a substantial, diverse, and representative sample of user preferences and interactions, making it an ideal resource for developing and testing recommendation algorithms.
		\item  \textbf{LastFM-1k}  a music recommendation dataset. Specifically, in preprocessing the LastFM-1k dataset, songs that have been listened to multiple times by each user are de-emphasized and songs that only have a play count of one across the item set are eliminated.
		\item  \textbf{Book-Crossing} a book recommendation dataset collected from the Book-Crossing community, where users rate and share books. In preprocessing, books with very few ratings (e.g., only one rating across all users) are typically removed to reduce noise. Additionally, inactive users with minimal interactions are often filtered out to ensure meaningful recommendation patterns. This preprocessing enhances the dataset's quality for training recommendation systems.
		
		\item  \textbf{Cold-Rec} a benchmark dataset designed for evaluating recommendation systems under cold-start scenarios. It contains both source and target datasets.	The target datasets, collected from Kandian recommender system over a one-month period, are utilized to evaluate accuracy via recommendation lists. It encompasses various interaction types, including news articles, videos, and advertisements.

	\end{itemize}

	\subsubsection{Models}\label{models}
	 Considering the absence of models specifically designed for user attribute classification and analyst identities in Scenarios 1 and 2, classical classification models are selected to predict user attributes:
	\begin{itemize}
		\item \textbf{DT} \cite{dt2} DT is characterized by ability to model complex decision-making processes through a hierarchical tree-like structure. DT perform feature selection by choosing the most significant feature at each node. 
		
		\item \textbf{KNN} \cite{knn} KNN is a simple yet effective algorithm used for classification by finding the K most similar data points to a given query and determining the output based on their majority or average value.
		
		\item \textbf{DNN} \cite{dnn} DNN is multi-layered neural networks capable of learning and representing highly complex patterns in data. Key features include its capacity for automatic feature extraction, robustness to noise, etc. 
		
		\item \textbf{PeterRec} \cite{peterrec} PeterRec transfers the knowledge from source domain to target domain based on transfer learning to mine a single user presentation, thereby predicting user profile.
	\end{itemize}
	
	In Scenarios 3 and 4, as the original recommendation model is generally a black box for analyst, who aims at selecting a model that has similar recommendation mechanism as the original one, thereby obtaining more useful information to analyze user behavior. To achieve such target, three widely used algorithms are chosen as the candidate recommendation algorithms of the spurious model: Matrix factorization based recommendation (MF); Neural network based recommendation (NeuMF); Graph based recommendation (NGCF). These three kinds of recommendation model could be on behalf of the most kinds of recommendation mechanism.	
    \begin{itemize}			
		\item \textbf{MF} \cite{mf0} MF is a collaborative filtering technique that uncovers latent features or factors in user-item interactions. It represents users and items in a lower-dimensional space, facilitating the prediction of missing entries.
		
		\item \textbf{NGCF} \cite{ngcf} NGCF has ability to model complex user-item relationships and to provide interpretable recommendations by learning from the local neighborhood structure of the graph. 
		
		\item \textbf{NeuMF} \cite{neumf} NeuMF leverages the linearity of MF to capture the latent factors in user-item interactions and the non-linearity of a Multi-Layer Perceptron (MLP) to learn complex patterns and representations from the data.
		
	\end{itemize}
	
	RAPI framework leverages pre-trained models to distill item content embedding, subsequently converting them into recommendation embedding by alignment model. The pre-trained models involved in the experiments are as follows:
	\begin{itemize}
		\item \textbf{BERT} \cite{bert} BERT is pre-trained on a large corpus, enabling it to understand complex linguistic patterns. Its multi-layer transformer architecture provides advanced capabilities for natural language processing tasks. BERT is versatile, serving for various tasks like sentiment analysis, question answering, and text classification.
		\item \textbf{gte-base} \cite{gte} 
		gte-base is a robust sentence embedding model optimized for tasks like information retrieval and semantic similarity. It leverages multi-stage contrastive learning for high-quality text representation.
		\item \textbf{gtr-t5-base} \cite{gtr} gtr-t5-base is a sentence transformer model that excels in semantic search tasks, converting sentences and paragraphs into a 768-dimensional dense vector space. It is optimized for retrieval tasks and benefits from the encoder of the t5-base model, which is known for its effectiveness in various NLP tasks. 
	\end{itemize}                  
	
	\subsubsection{Baselines} 

	  To assess potential problems in recommendation lists, RAPI framework is introduced to infer users' sensitive attributes, with scenario 1 and 2 serving as baseline comparisons in its experimental setup. Specially, aforementioned models (e.g., KNN) serve as classification models to infer user attributes in scenario 1 and 2, respectively. 
	  Furthermore, an advanced method, PeterRec\cite{peterrec}, is also employed as a comparative baseline in scenario 2, which leverages transfer learning to mine individual user representations for user profile prediction.

	\subsubsection{Parameter Settings}
	RAPI framework is implemented with Pytorch.
	By default, the original recommendation model is LightGCN \cite{lightgcn}, which is a simplified graph convolutional network (GCN) model designed for efficient collaborative filtering in recommendation systems. The analyst obtains top-20 recommendation lists, and the data augmented recommendation list is top-50. The recommendation embedding for both the original recommendation model and the spurious model is 64 dimensions, and the content embedding from BERT is 768 dimensions.
	All model training procedures
	are facilitated using the Adam optimizer, employing a learning rate
	searched in [0.05, 0.01, 0.005, 0.001]. The batch size is searched in [128, 256, 512, 1024]. The default training epochs is 500, early stop technique is utilized to prevent overfitting. 
	
    \subsubsection{Statistics about Sensitive Attributes.}
	
	\begin{figure}
		\centering
		\begin{subfigure}{0.48\linewidth}
			\includegraphics[width=\linewidth]{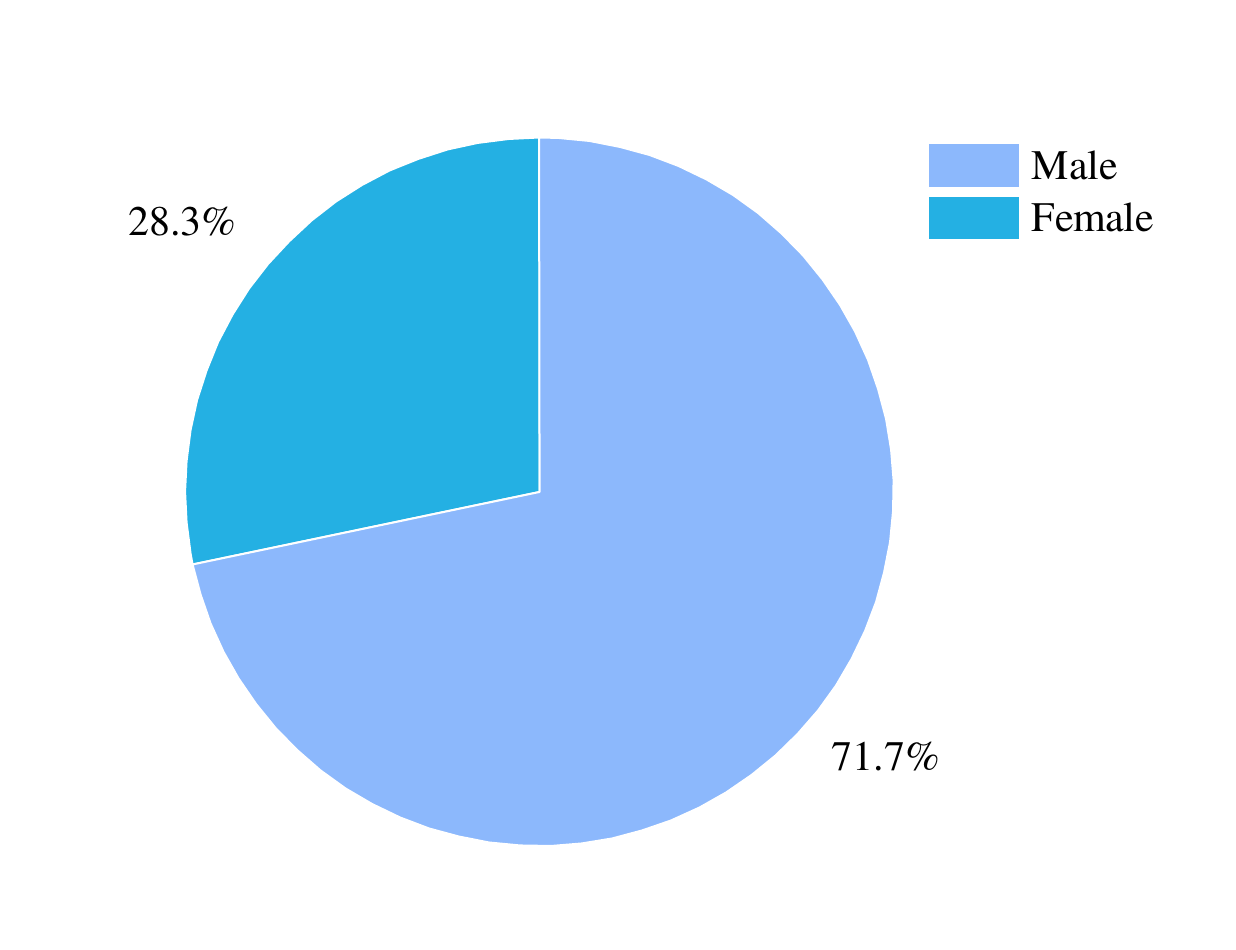}
			\caption{Gender distribution in MovieLens-1m.}
		\end{subfigure}
		\begin{subfigure}{0.48\linewidth}
			\includegraphics[width=\linewidth]{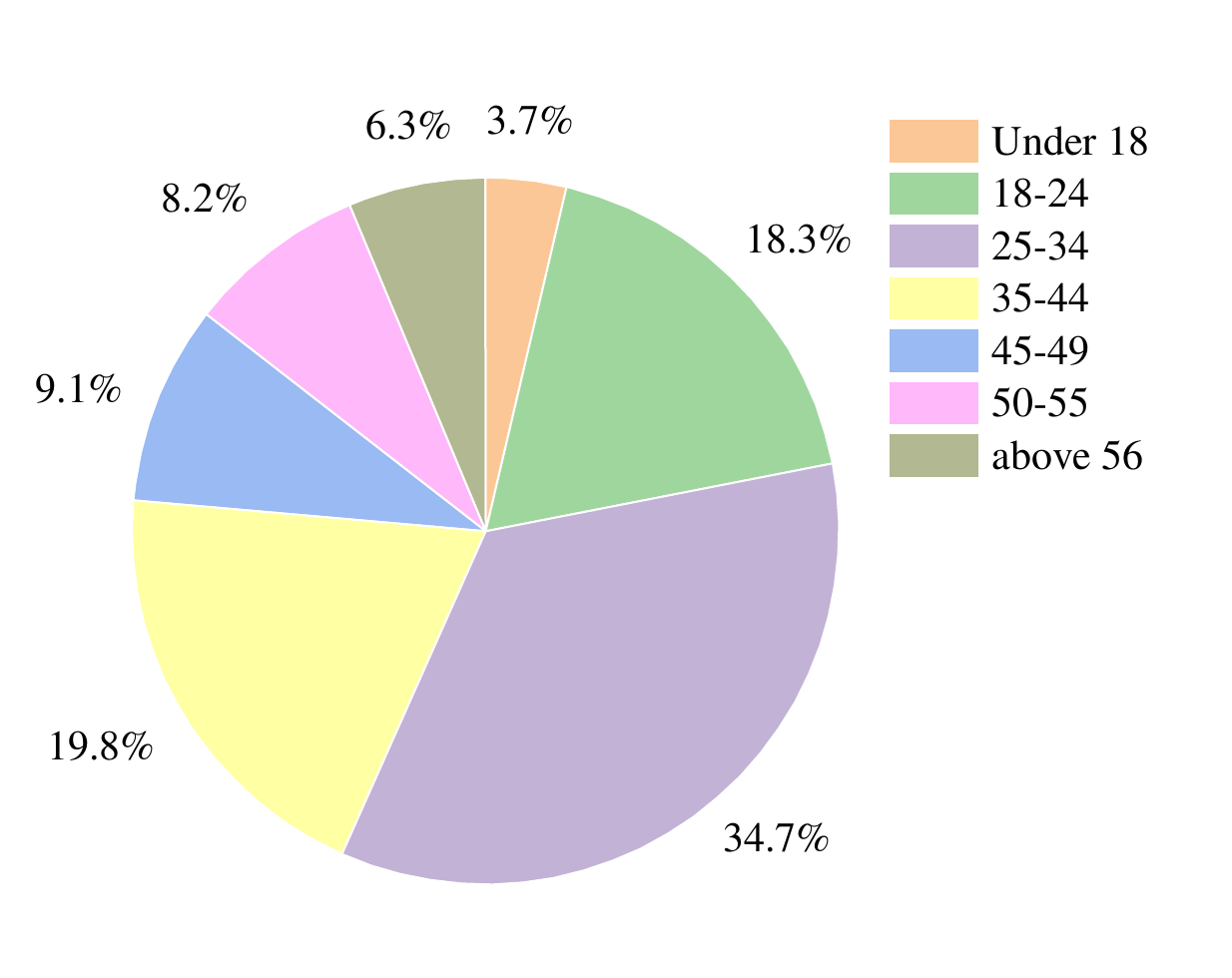}
			\caption{ Age distribution in MovieLens-1m.}
		\end{subfigure}

			\begin{subfigure}{0.48\linewidth}
			\includegraphics[width=\linewidth]{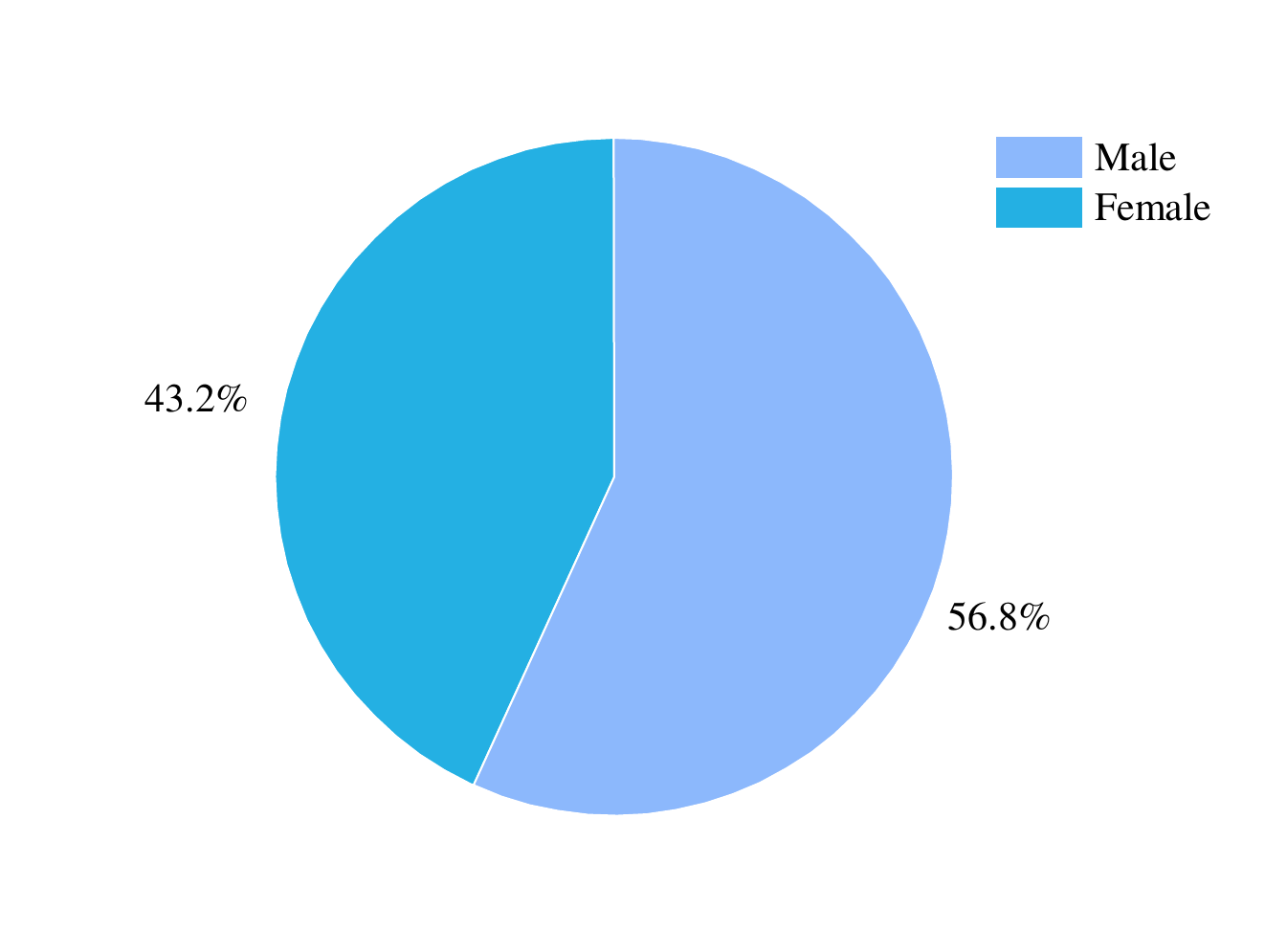}
			\caption{ Gender distribution in LastFM-1k.}
		\end{subfigure}
		\begin{subfigure}{0.48\linewidth}
		\includegraphics[width=\linewidth]{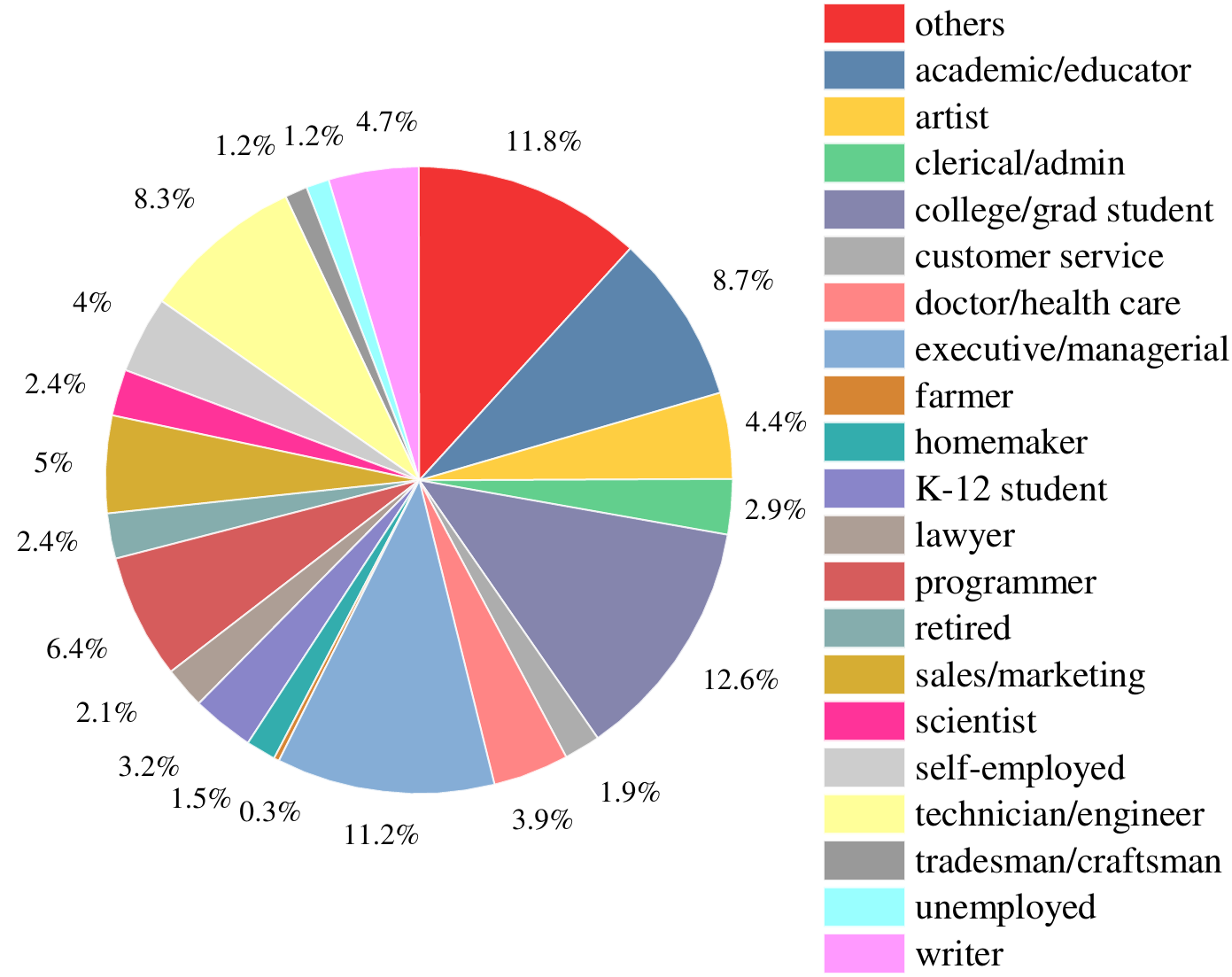}
		\caption{ Occupation distribution in MovieLens-1m.}
		\end{subfigure}

		\caption{Statistics about sensitive attributes in records. }
		\label{statistics}
	\end{figure}
	
	Detailed statistics about sensitive attributes in MovieLens-1m and LastFM-1k datasets are shown in figure \ref{statistics}. This will facilitate understanding of the distribution and characteristics of data used in experiments and verify the effectiveness of RAPI framework. Note that users with blank gender in the LastFM-1k datasets are excluded from experiments.

	\subsection{ Performance Evaluation (\textbf{RQ1})}\label{RQ1}
	
\begin{table}[]
	\centering
	\caption{Inference results for gender of sensitive attributes across different scenarios in MovieLens-1m. (where $\alpha$, $\beta$ represent the proportion of interaction-released providers and the proportion of sensitive attribute-released providers, respectively.)}
	\begin{tabular}{c|c|cccccc}
		\hline
		\multirow{2}{*}{scenario} & \multirow{2}{*}{method} & \multicolumn{6}{c}{MovieLens-1m}                                                                 \\ \cline{3-8}
		&                         & \multicolumn{1}{c|}{$\alpha$}   & acc($\beta$=0.1) & acc($\beta$=0.3) & acc($\beta$=0.5) & acc($\beta$=0.7) & acc($\beta$=0.9) \\ \hline
		\multirow{3}{*}{1}        & DT                      & \multicolumn{1}{c|}{0}   & 0.5958     & 0.5999     & 0.6021    & 0.6038     & 0.6041     \\
		& KNN                     & \multicolumn{1}{c|}{0}   & 0.6170    & 0.6175    & 0.6248    & 0.6447     & 0.6497     \\
		& DNN                     & \multicolumn{1}{c|}{0}   & 0.6801     & 0.6976     & 0.7014     & 0.7086     & 0.7103     \\ \hline
		\multirow{4}{*}{2}        & DT                      & \multicolumn{1}{c|}{0}   & 0.6549     & 0.6560     & 0.6559     & 0.6634     & 0.6904     \\
		& KNN                     & \multicolumn{1}{c|}{0}   & 0.6921     & 0.6968     & 0.6974     & 0.6981     & 0.6999     \\
		& DNN                     & \multicolumn{1}{c|}{0}   & 0.7316     & 0.7336     & 0.7342     & 0.7355     & 0.7378     \\ 
			& PeterRec                     & \multicolumn{1}{c|}{0}   & 0.7331     & 0.7346     & 0.7348     & 0.7361     & 0.7381     \\\hline
		3                         & RAPI                     & \multicolumn{1}{c|}{0}   & 0.7345     & 0.7358   & 0.7368     & 0.7379     & 0.7390    \\ \hline
		\multirow{5}{*}{4}        & \multirow{5}{*}{RAPI}    & \multicolumn{1}{c|}{0.1} & 0.7517     & 0.7529     & 0.7556     & 0.7557     & 0.7585    \\
		&                         & \multicolumn{1}{c|}{0.3} & 0.7539     & 0.7546     & 0.7550    & 0.7568    & 0.7569     \\
		&                         & \multicolumn{1}{c|}{0.5} & 0.7541     & 0.7548     & 0.7552    & 0.7552     & 0.7586     \\
		&                         & \multicolumn{1}{c|}{0.7} & 0.7543    & 0.7556 & 0.7558     & 0.7569    & 0.7570     \\
		&                         & \multicolumn{1}{c|}{0.9} & 0.7614     & 0.7623      & 0.7631     & 0.7634     & 0.7640     \\ \hline
	\end{tabular}
	\label{t1}
\end{table}

	\begin{table}[]
		\centering
		\caption{Inference results for gender of sensitive attributes across different scenarios in LastFM-1k.}
		\begin{tabular}{c|c|cccccc}
			\hline
			\multirow{2}{*}{scenario} & \multirow{2}{*}{method} & \multicolumn{6}{c}{LastFM-1k}                                                                 \\ \cline{3-8}
			&                         & \multicolumn{1}{c|}{$\alpha$}   & acc($\beta$=0.1) & acc($\beta$=0.3) & acc($\beta$=0.5) & acc($\beta$=0.7) & acc($\beta$=0.9) \\ \hline
			\multirow{3}{*}{1}        & DT                      & \multicolumn{1}{c|}{0}   & 0.4976     & 0.5001     & 0.5038    & 0.5100    & 0.5316     \\
			& KNN                     & \multicolumn{1}{c|}{0}   & 0.4976    & 0.5294    & 0.5314    & 0.5338     & 0.5506     \\
			& DNN                     & \multicolumn{1}{c|}{0}   & 0.5089     & 0.5169     & 0.5197     & 0.5271     & 0.5677     \\ \hline
			\multirow{4}{*}{2}        & DT                      & \multicolumn{1}{c|}{0}   & 0.5188     & 0.5541    & 0.5656     & 0.5789     & 0.5955     \\
			& KNN                     & \multicolumn{1}{c|}{0}   & 0.5698     & 0.5699     & 0.5815     & 0.5818     & 0.5849     \\
			& DNN                     & \multicolumn{1}{c|}{0}   & 0.5723     & 0.5725     & 0.5928     & 0.6102     & 0.6164     \\
			& PeterRec                     & \multicolumn{1}{c|}{0}   & 0.5721     & 0.5736     & 0.5932     & 0.6124     & 0.6169     \\ \hline
			3                         & RAPI                     & \multicolumn{1}{c|}{0}   & 0.5723     & 0.5809    & 0.5951     & 0.6136     & 0.6226    \\ \hline
			\multirow{5}{*}{4}        & \multirow{5}{*}{RAPI}    & \multicolumn{1}{c|}{0.1} & 0.5723    & 0.5792     & 0.5962     & 0.6137     & 0.6233    \\
			&                         & \multicolumn{1}{c|}{0.3} & 0.5723     & 0.5809     & 0.5959    & 0.6136    & 0.6237     \\
			&                         & \multicolumn{1}{c|}{0.5} & 0.5736     & 0.5811 & 0.5962    & 0.6141     & 0.6241     \\
			&                         & \multicolumn{1}{c|}{0.7} & 0.5739    & 0.5814 & 0.5973     & 0.6151    & 0.6268     \\
			&                         & \multicolumn{1}{c|}{0.9} & 0.5742     & 0.5821      & 0.5995     & 0.6157     & 0.6477    \\ \hline
		\end{tabular}
		\label{t2}
	\end{table}
	
	\begin{table}[]
		\centering
		\caption{Inference results for age of sensitive attributes across different scenarios in MovieLens-1m.}
		\begin{tabular}{c|c|cccccc}
				\hline
				\multirow{2}{*}{scenario} & \multirow{2}{*}{method} & \multicolumn{6}{c}{MovieLens-1m}                                                                 \\ \cline{3-8}
				&                         & \multicolumn{1}{c|}{$\alpha$}   & acc($\beta$=0.1) & acc($\beta$=0.3) & acc($\beta$=0.5) & acc($\beta$=0.7) & acc($\beta$=0.9) \\ \hline
				\multirow{3}{*}{1}        & DT                      & \multicolumn{1}{c|}{0}   & 0.2211     & 0.2301     & 0.2268     & 0.2213     & 0.2202     \\
				& KNN                     & \multicolumn{1}{c|}{0}   & 0.2371     & 0.2226     & 0.2212     & 0.2328     & 0.2219     \\
				& DNN                     & \multicolumn{1}{c|}{0}   & 0.3269     & 0.3463     & 0.3497     & 0.3460      & 0.3361     \\ \hline
				\multirow{4}{*}{2}        & DT                      & \multicolumn{1}{c|}{0}   & 0.2662     & 0.2675     & 0.2566     & 0.2864     & 0.2583     \\
				& KNN                     & \multicolumn{1}{c|}{0}   & 0.2997     & 0.2942     & 0.3003     & 0.3135     & 0.2913     \\
				& DNN                     & \multicolumn{1}{c|}{0}   & 0.3441     & 0.3576     & 0.3672     & 0.3576     & 0.3493     \\
				& PeterRec                    & \multicolumn{1}{c|}{0}   & 0.3469     & 0.3579     & 0.3713     & 0.3729     & 0.3731     \\ \hline
				3                         & RAPI                     & \multicolumn{1}{c|}{0}   & 0.3510      & 0.3595     & 0.3828     & 0.3838     & 0.4123     \\ \hline
				\multirow{5}{*}{4}        & \multirow{5}{*}{RAPI}    & \multicolumn{1}{c|}{0.1} & 0.3714     & 0.3881     & 0.4291     & 0.4143     & 0.4454    \\
				&                         & \multicolumn{1}{c|}{0.3} & 0.3769     & 0.3855     & 0.4285     & 0.4183     & 0.4487     \\
				&                         & \multicolumn{1}{c|}{0.5} & 0.3859     & 0.3964     & 0.4285     & 0.4289     & 0.4553     \\
				&                         & \multicolumn{1}{c|}{0.7} & 0.3859     & 0.4059     & 0.4311     & 0.4245     & 0.4555     \\
				&                         & \multicolumn{1}{c|}{0.9} & 0.3883     & 0.4130      & 0.4421     & 0.4387     & 0.4652     \\ \hline
		\end{tabular}
		\label{t3}
	\end{table}
	
	\begin{table}[]
		\centering
		\caption{Inference results for occupation of sensitive attributes across different scenarios in MovieLens-1m.}
		\begin{tabular}{c|c|cccccc}
				\hline
				\multirow{2}{*}{scenario} & \multirow{2}{*}{method} & \multicolumn{6}{c}{MovieLens-1m}                                                                 \\ \cline{3-8}
				&                         & \multicolumn{1}{c|}{$\alpha$}   & acc($\beta$=0.1) & acc($\beta$=0.3) & acc($\beta$=0.5) & acc($\beta$=0.7) & acc($\beta$=0.9) \\ \hline
				\multirow{3}{*}{1}        & DT                      & \multicolumn{1}{c|}{0}   & 0.0793     & 0.0788     & 0.0805     & 0.0778     & 0.0811     \\
				& KNN                     & \multicolumn{1}{c|}{0}   & 0.0760      & 0.0766     & 0.0778     & 0.0728     & 0.0778     \\
				& DNN                     & \multicolumn{1}{c|}{0}   & 0.1255     & 0.1185     & 0.1182     & 0.1230      & 0.1275     \\ \hline
				\multirow{4}{*}{2}        & DT                      & \multicolumn{1}{c|}{0}   & 0.0887     & 0.0889     & 0.0848     & 0.0922     & 0.0811     \\
				& KNN                     & \multicolumn{1}{c|}{0}   & 0.0885     & 0.0937     & 0.0964     & 0.1015     & 0.0911     \\
				& DNN                     & \multicolumn{1}{c|}{0}   & 0.1277     & 0.1362     & 0.1430     & 0.1363     & 0.1672     \\
				& PeterRec                    & \multicolumn{1}{c|}{0}   & 0.1251     & 0.1262     & 0.1266     & 0.1365     & 0.1569     \\ \hline
				3                         & RAPI                     & \multicolumn{1}{c|}{0}   & 0.1262     & 0.1272     & 0.1321     & 0.1485     & 0.1672     \\ \hline
				\multirow{5}{*}{4}        & \multirow{5}{*}{RAPI}    & \multicolumn{1}{c|}{0.1} & 0.1543     & 0.1670      & 0.1702     & 0.1893     & 0.2003     \\
				&                         & \multicolumn{1}{c|}{0.3} & 0.1538     & 0.1686     & 0.1743     & 0.1887     & 0.2036     \\
				&                         & \multicolumn{1}{c|}{0.5} & 0.1556     & 0.1686     & 0.1762     & 0.1766     & 0.2119     \\
				&                         & \multicolumn{1}{c|}{0.7} & 0.1562     & 0.1708     & 0.1762     & 0.1755     & 0.2136     \\
				&                         & \multicolumn{1}{c|}{0.9} & 0.1551     & 0.1746     & 0.1791     & 0.1893     & 0.2137     \\ \hline
		\end{tabular}
		\label{t4}
	\end{table}
	
		\begin{table}[]
		\centering
		\caption{Inference results for age of sensitive attributes across different scenarios in Book-Crossing.}
		\begin{tabular}{c|c|cccccc}
			\hline
			\multirow{2}{*}{scenario} & \multirow{2}{*}{method} & \multicolumn{6}{c}{Book-Crossing}                                                                 \\ \cline{3-8}
			&                         & \multicolumn{1}{c|}{$\alpha$}   & acc($\beta$=0.1) & acc($\beta$=0.3) & acc($\beta$=0.5) & acc($\beta$=0.7) & acc($\beta$=0.9) \\ \hline
			\multirow{3}{*}{1}        & DT                      & \multicolumn{1}{c|}{0}   & 0.1529     & 0.1535     & 0.1541     & 0.1544     & 0.1551     \\
			& KNN                     & \multicolumn{1}{c|}{0}   & 0.1553      & 0.1561     & 0.1563     & 0.1571     & 0.1572     \\
			& DNN                     & \multicolumn{1}{c|}{0}   & 0.1779     & 0.1781     & 0.1786     & 0.1789      & 0.1796     \\ \hline
			\multirow{4}{*}{2}        & DT                      & \multicolumn{1}{c|}{0}   & 0.1878     & 0.1889     & 0.1891     & 0.1897     & 0.1899     \\
			& KNN                     & \multicolumn{1}{c|}{0}   & 0.1911     & 0.1919     & 0.1923     & 0.1926     & 0.1928     \\
			& DNN                     & \multicolumn{1}{c|}{0}   & 0.1932     & 0.1934     & 0.1941     & 0.1948     & 0.1952     \\
			& PeterRec                    & \multicolumn{1}{c|}{0}   & 0.1996     & 0.1999     & 0.2011     & 0.2019     & 0.2022     \\ \hline
			3                         & RAPI                     & \multicolumn{1}{c|}{0}   & 0.2123     & 0.2142     & 0.2158     & 0.2169     & 0.2173     \\ \hline
			\multirow{5}{*}{4}        & \multirow{5}{*}{RAPI}    & \multicolumn{1}{c|}{0.1} & 0.2346     & 0.2349      & 0.2355     & 0.2364     & 0.2367     \\
			&                         & \multicolumn{1}{c|}{0.3} & 0.2351     & 0.2356     & 0.2359     & 0.2362     & 0.2369     \\
			&                         & \multicolumn{1}{c|}{0.5} & 0.2358     & 0.2360    & 0.2365     & 0.2367     & 0.2368    \\
			&                         & \multicolumn{1}{c|}{0.7} & 0.2364     & 0.2366     & 0.2368     & 0.2371     & 0.2376     \\
			&                         & \multicolumn{1}{c|}{0.9} & 0.2377     & 0.2378     & 0.2379    & 0.2383     & 0.2385     \\ \hline
		\end{tabular}
		\label{t5}
	\end{table}
	
	\begin{table}[]
		\centering
		\caption{Inference results for gender of sensitive attributes across different scenarios in Cold-Rec.}
		\begin{tabular}{c|c|cccccc}
			\hline
			\multirow{2}{*}{scenario} & \multirow{2}{*}{method} & \multicolumn{6}{c}{Cold-Rec}                                                                 \\ \cline{3-8}
			&                         & \multicolumn{1}{c|}{$\alpha$}   & acc($\beta$=0.1) & acc($\beta$=0.3) & acc($\beta$=0.5) & acc($\beta$=0.7) & acc($\beta$=0.9) \\ \hline
			\multirow{3}{*}{1}        & DT                      & \multicolumn{1}{c|}{0}   & 0.5134     & 0.5136     & 0.5139     & 0.5142     & 0.5146     \\
			& KNN                     & \multicolumn{1}{c|}{0}   & 0.5278      & 0.5283     & 0.5288     & 0.5289     & 0.5291     \\
			& DNN                     & \multicolumn{1}{c|}{0}   & 0.5667     & 0.5673     & 0.5683     & 0.5689      & 0.5694     \\ \hline
			\multirow{4}{*}{2}        & DT                      & \multicolumn{1}{c|}{0}   & 0.5612     & 0.5621     & 0.5629     & 0.5633     & 0.5641     \\
			& KNN                     & \multicolumn{1}{c|}{0}   & 0.5627     & 0.5628     & 0.5631     & 0.5633     & 0.5637     \\
			& DNN                     & \multicolumn{1}{c|}{0}   & 0.5911     & 0.5924     & 0.5931     & 0.5942     & 0.5949     \\
			& PeterRec                    & \multicolumn{1}{c|}{0}   & 0.6636     & 0.6638     & 0.6649     & 0.6651     & 0.6659     \\ \hline
			3                         & RAPI                     & \multicolumn{1}{c|}{0}   & 0.6711     & 0.6712     & 0.6719     & 0.6724     & 0.6725     \\ \hline
			\multirow{5}{*}{4}        & \multirow{5}{*}{RAPI}    & \multicolumn{1}{c|}{0.1} & 0.6725     & 0.6728      & 0.6729     & 0.6731     & 0.6732     \\
			&                         & \multicolumn{1}{c|}{0.3} & 0.6727     & 0.6729     & 0.6731     & 0.6734     & 0.6742     \\
			&                         & \multicolumn{1}{c|}{0.5} & 0.6731    & 0.6734    & 0.6738     & 0.6741     & 0.6744    \\
			&                         & \multicolumn{1}{c|}{0.7} & 0.6735     & 0.6737     & 0.6741     & 0.6743     & 0.6747     \\
			&                         & \multicolumn{1}{c|}{0.9} & 0.6739     & 0.6741     & 0.6748    & 0.6751     & 0.6756     \\ \hline
		\end{tabular}
		\label{t6}
	\end{table}

	The experimental results are shown in \cref{t1,t2,t3,t4,t5,t6}. In order to assess the probability of user information exposure comprehensively, the information providers are divided into two parts: one part is providers who release interaction records, and the other part is providers who release sensitive attributes. The proportion of users accounted for by the former is denoted by $\alpha$, and the latter is denoted by $\beta$. The evaluation metrics of gender, age, and occupation are evaluated in MovieLens-1m, while gender is assessed in LastFM-1k and Cold-Rec, and age is analyzed in Book-Crossing, as shown in table \ref{dataset}.
	\par
	Inference of gender, age, and occupation can achieve varying levels of accuracy for analysts in different scenarios, treated as a binary, seven-class, and twenty-one-class classification problem, respectively. This suggests that there is indeed an information exposure concern associated with user's recommendation list potentially revealing the user's personal profiles. Specifically, in the MovieLens-1m information set, the highest classification accuracies for gender, age, and occupation inference are 0.7640, 0.4652, and 0.2137, respectively. For the LastFM-1k information set, gender inference achieves an accuracy of 0.6477. Additionally, the highest inference accuracies for age and gender in the Book-Crossing and Cold-Rec information sets are 0.2385 and 0.6756, respectively.
	\par
	Furthermore, when considering different sensitive attributes, it is observed that scenario 1 exhibits the lowest classification accuracy, while scenarios 2 and 3 have comparable accuracies, and scenario 4 achieves the highest classification accuracy. This trend is logical, given that in Scenario 1, the analyst relies solely on the item number to represent the item, while the analyst can utilize the item's recommendation embedding additionally in scenario 2. Scenario 3 involves the use of the item's content embedding, and in Scenario 4, the analyst leverages both the item's recommendation embedding and content embedding. Scenarios 3 and 4 demonstrate that incorporating content embedding as supplementary information significantly enhances inference accuracy.
	\par
	What's more, when considering different scenarios within specific sensitive attribute, e.g. gender in MovieLens-1m, following observations suggest that RAPI is more capable of mining attribute representations than existing advanced methods and can be further developed as more information becomes available:
	\begin{enumerate}
		\item Compared with DT and KNN, DNN is more suitable to infer  attribute with higher accuracy due to its capacity of strong representation mining in scenarios 1 and 2.
		\item Leveraging knowledge transfer from the source domain, PeterRec achieves superior performance in scenario 2 by generating user representations for profile prediction through sequential interaction modeling.
		\item The RAPI shows a slightly superior performance compared to DNN and PeterRec in scenario 3, while demonstrating enhanced inference capability with increasing $\beta$ values in scenario 4.
	\end{enumerate}
	
	\par
	With respect to scenarios 3 and 4 of the RAPI framework, the influence of the two hyperparameters $\alpha$ and $\beta$ on RAPI are discussed. $\alpha$ and $\beta$ represent the proportion of interaction-released providers and sensitive attribute-released providers, respectively. As $\alpha$ and $\beta$ increases, the classification accuracy usually improves slowly. The reason is that with the higher number of providers, the spurious model can obtain more recommendation embedding of provider-related items, and subsequently the transformed model can generate more accurate recommendation embedding; the more the number of users with known sensitive attributes, the more supervised information can be obtained by the adaptive weight classification model, which effectively improves the inference accuracy.
	\par
   Experimental results on the Book Crossing and Cold Rec datasets validate the scalability and effectiveness of the RAPI framework. Across diverse recommendation domains including movies, music, books, and news, RAPI consistently outperforms existing methods in inferring user attributes via recommendation lists.

   Given the distribution imbalance existing in the datasets, relying solely on the accuracy may not be able to comprehensively evaluate the overall performance of RAPI. The table \ref{t7} reports F1-score metrics to enable a more robust and unbiased evaluation of inference performance.

	\begin{table}[]
		\centering
		\caption{F1-score presentation when inferring gender attributes in MovieLens-1m.}
		\begin{tabular}{c|c|cccccc}
			\hline
			\multirow{2}{*}{scenario} & \multirow{2}{*}{method} & \multicolumn{6}{c}{MovieLens-1m}                                                                 \\ \cline{3-8}
			&                         & \multicolumn{1}{c|}{$\alpha$}   & F1 ($\beta$=0.1) & F1 ($\beta$=0.3) & F1 ($\beta$=0.5) & F1 ($\beta$=0.7) & F1 ($\beta$=0.9) \\ \hline
			\multirow{3}{*}{1}        & DT                      & \multicolumn{1}{c|}{0}   & 0.5111     & 0.5112     & 0.5114    & 0.5116     & 0.5117     \\
			& KNN                     & \multicolumn{1}{c|}{0}   & 0.5673    & 0.5678   & 0.5679    & 0.5679     & 0.5681     \\
			& DNN                     & \multicolumn{1}{c|}{0}   & 0.6413     & 0.6416     & 0.6419     & 0.6422     & 0.6425     \\ \hline
			\multirow{4}{*}{2}        & DT                      & \multicolumn{1}{c|}{0}   & 0.5677     & 0.5679     & 0.5681     & 0.5682     & 0.5685     \\
			& KNN                     & \multicolumn{1}{c|}{0}   & 0.6312    & 0.6314     & 0.6317     & 0.6319    & 0.6321     \\
			& DNN                     & \multicolumn{1}{c|}{0}   & 0.6441     & 0.6443     & 0.6444     & 0.6445     & 0.6448     \\ 
			& PeterRec                     & \multicolumn{1}{c|}{0}   & 0.6456     & 0.6457     & 0.6457     & 0.6459     & 0.6451     \\\hline
			3                         & RAPI                     & \multicolumn{1}{c|}{0}   & 0.6495     & 0.6497    & 0.6498     & 0.6501     & 0.6504    \\ \hline
			\multirow{5}{*}{4}        & \multirow{5}{*}{RAPI}    & \multicolumn{1}{c|}{0.1} & 0.6651     & 0.6654    & 0.6656     & 0.6658     & 0.6661    \\
			&                         & \multicolumn{1}{c|}{0.3} & 0.6655     & 0.6656     & 0.6658    & 0.6661    & 0.6664     \\
			&                         & \multicolumn{1}{c|}{0.5} & 0.6659     & 0.6662     & 0.6664    & 0.6668     & 0.6671     \\
			&                         & \multicolumn{1}{c|}{0.7} & 0.6659    & 0.6661 & 0.6663     & 0.6668    & 0.6672     \\
			&                         & \multicolumn{1}{c|}{0.9} & 0.6662     & 0.6664      & 0.6667     & 0.6667     & 0.6669     \\ \hline
		\end{tabular}
		\label{t7}
	\end{table}
	
	\subsection{Impact of Various Modules within RAPI Framework (\textbf{RQ2})}
	In this section, impact of various modules within RAPI framework (e.g. spurious model confirmation, data augmentation for recommendation list and adaptive weight classification model) is investigated respectively.
	
	\begin{figure*}
		\centering
		\includegraphics[width=\linewidth]{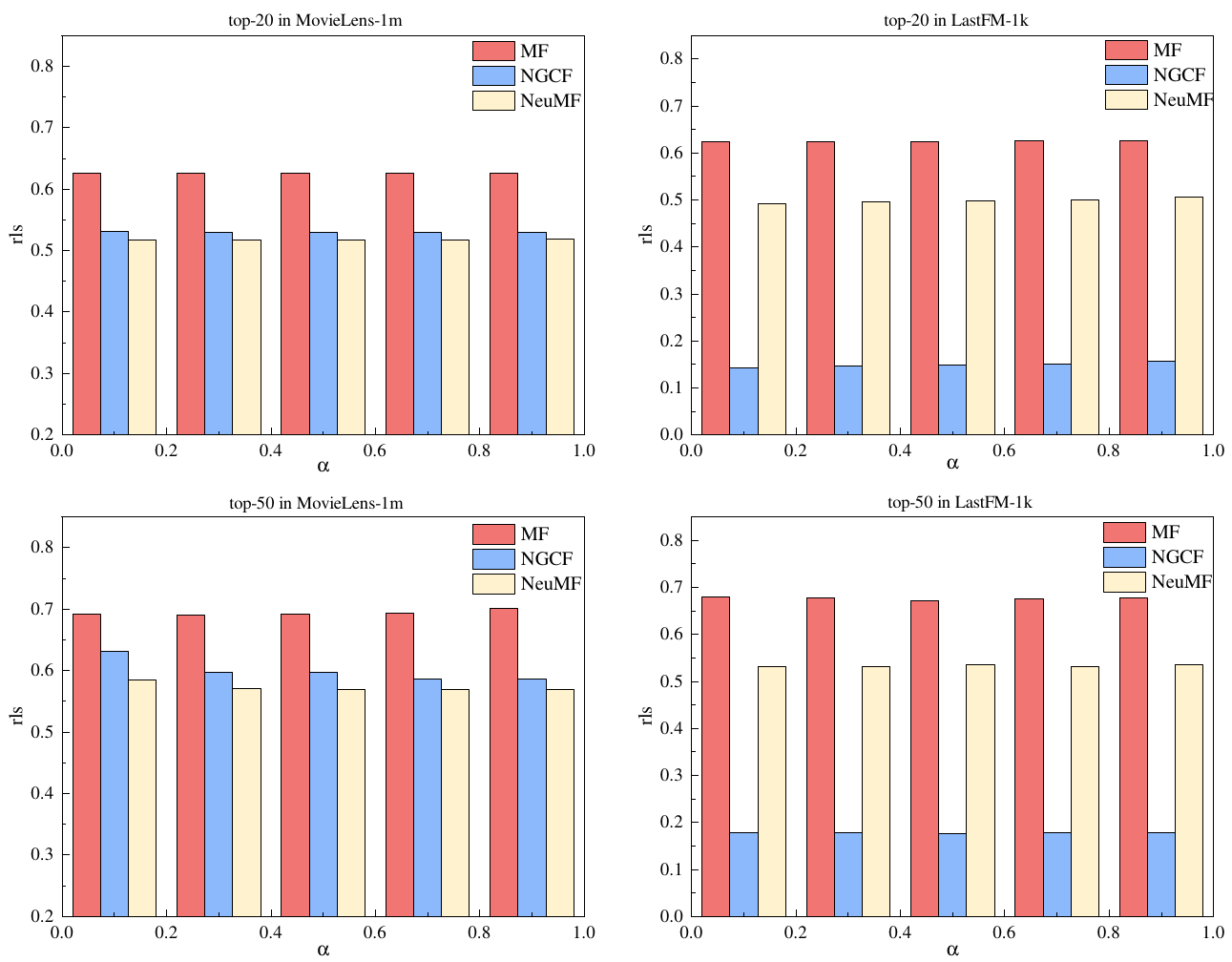}
		\caption{Recommendation performance of various candidate models.}
		\label{reclist}
	\end{figure*}

	\subsubsection{Recommendation list similarity in spurious model confirmation}
	
	To construct comprehensive user profiles as accurately as possible, recommendation lists from the original recommendation model are compared with those from candidate models to confirm an optimal profiling model. The experimental results are shown in Figure \ref{reclist}, where the $\mathit{rls}$ defined by equation \ref{rls} is the percentage of recommended items hitting the original recommendation list. As discussed in section \ref{models}, the profiling model is selected from three classical recommendation models: matrix factorization based recommendation (MF), neural network based recommendation (NeuMF) and graph based recommendation (NGCF). Several observations are summarized as follows:
	\begin{enumerate}
		\item MF achieves the most similar recommendation performance to LightGCN on datasets. Specifically, the similarity between the top-20, top-50 recommendation lists generated by MF and their LightGCN counterparts is around 0.62 and 0.69 in MovieLens-1m, and around 0.62 and 0.68 in LastFM-1k, which means that MF is more suitable as a profiling model participating in subsequent procedures for constructing user profiles compared with NeuMF and NGCF.
		\item The similarity between recommendation lists generated by various candidate models and original recommendation lists is not affected by $\alpha$, this may be due to the fact that even if the percentage of users sharing interaction data $\alpha$ is extremely small, e.g. 0.1\%, it is possible to expose a large portion of the items, resulting in a subsequent increase in $\alpha$ with less impact on $rls$.
		\item The degree to which the profiling model's recommendation lists mirror those of the original recommendation model fluctuates with different datasets. Specifically, \(rls\) between NGCF and LightGCN is greater than that between NeuMF and LightGCN in the MovieLens-1m, whereas the situation is reversed in LastFM-1k.
	\end{enumerate}

	\subsubsection{Recommendation embedding similarity in data augmentation for recommendation list}
	The experimental results are presented in table \ref{r5}. The $\mathit {res}$ defined by equation \ref{res} quantifies the euclidean distance between the recommendation embedding derived from the content embedding of user-related items and the original recommendation embedding; a smaller $\mathit {res}$ indicates better performance of the alignment model. In order to ensure the generality of the alignment models within the proposed framework, two native yet effective models, MLP and AE, are selected as the candidate alignment models. Observations are summarized as follows:
	\begin{enumerate}
		\item Overall, MLP is more suitable than AE as an alignment model within the proposed framework. The \(/\) in table \ref{r5} signifies that there are no user-unrelated items in the LastFM-1k dataset when the proportion of data-sharing users is 0.7 or higher, implying that the profiling model has retained the recommendation embedding for all items.
		
        \item Recommendation embedding for items in the MovieLens-1m is approximately 4 units distant from the content embedding derived from the pre-trained BERT model, whereas the corresponding distance in the LastFM-1k is approximately 30 units. This implies that the recommendation embedding in the MovieLens-1m are richer in content information and, consequently, may be more susceptible to content-based analysis.
		
        \item Performance of alignment model which aims at transforming content embedding to recommendation embedding is affected by $\alpha$. The higher the value of $\alpha$, the better the alignment performance becomes, due to the fact that more recommendation embedding is available, and more supervisory information is involved in the training of alignment model.
	\end{enumerate}

	\begin{table*}[]
		\centering
		\caption{$\mathit {res}$ in transformation of content embedding from BERT model to recommendation embedding.}
		\begin{tabular}{c|ccccc|ccccc}
			\hline
			\multirow{2}{*}{method} & \multicolumn{5}{c|}{MovieLens-1m}                                     & \multicolumn{5}{c}{LastFM-1k}                                  \\ \cline{2-11}
			& $\alpha$=0.1 & $\alpha$=0.3 & $\alpha$=0.5 & $\alpha$=0.7 & $\alpha$=0.9 & $\alpha$=0.1 & $\alpha$=0.3 & $\alpha$=0.5 & $\alpha$=0.7 & $\alpha$=0.9 \\ \hline
			MLP                     &4.334     & 3.901    & 3.873     & 3.468     & 3.397     & 31.915    & 29.753    & 28.036    & /          & /          \\ \hline
			AE                      & 4.391     &4.179     & 3.958     & 3.931     & 3.470     &34.821    & 31.555    & 30.673    & /          & /          \\ \hline
		\end{tabular}
		\label{r5}
	\end{table*}
		
	\begin{figure}[]
		\centering
		\begin{subfigure}{0.49\linewidth}
			\includegraphics[width=\linewidth]{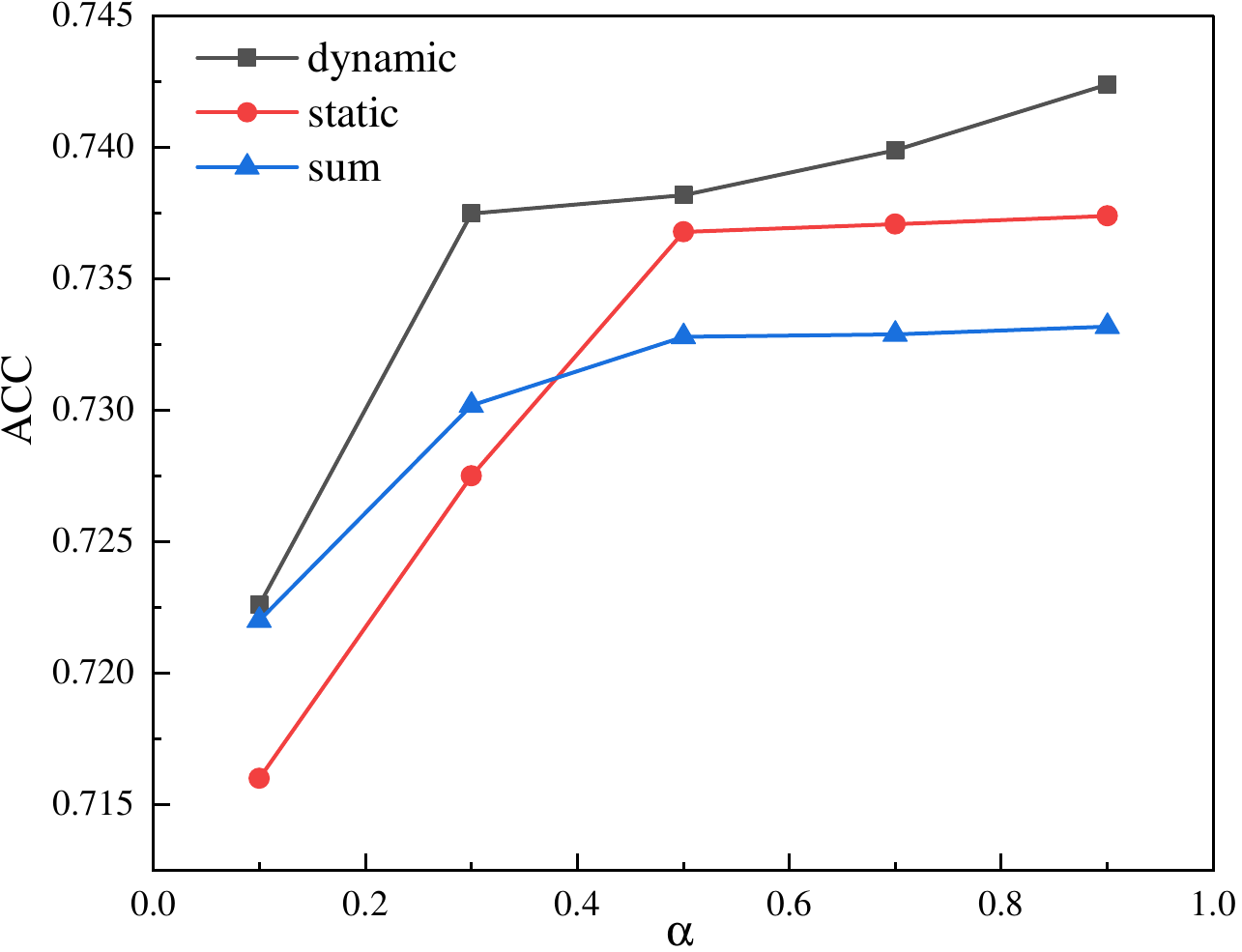}
			\caption{ MovieLens-1m}
		\end{subfigure}
		\hfill
		\begin{subfigure}{0.49\linewidth}
			\includegraphics[width=\linewidth]{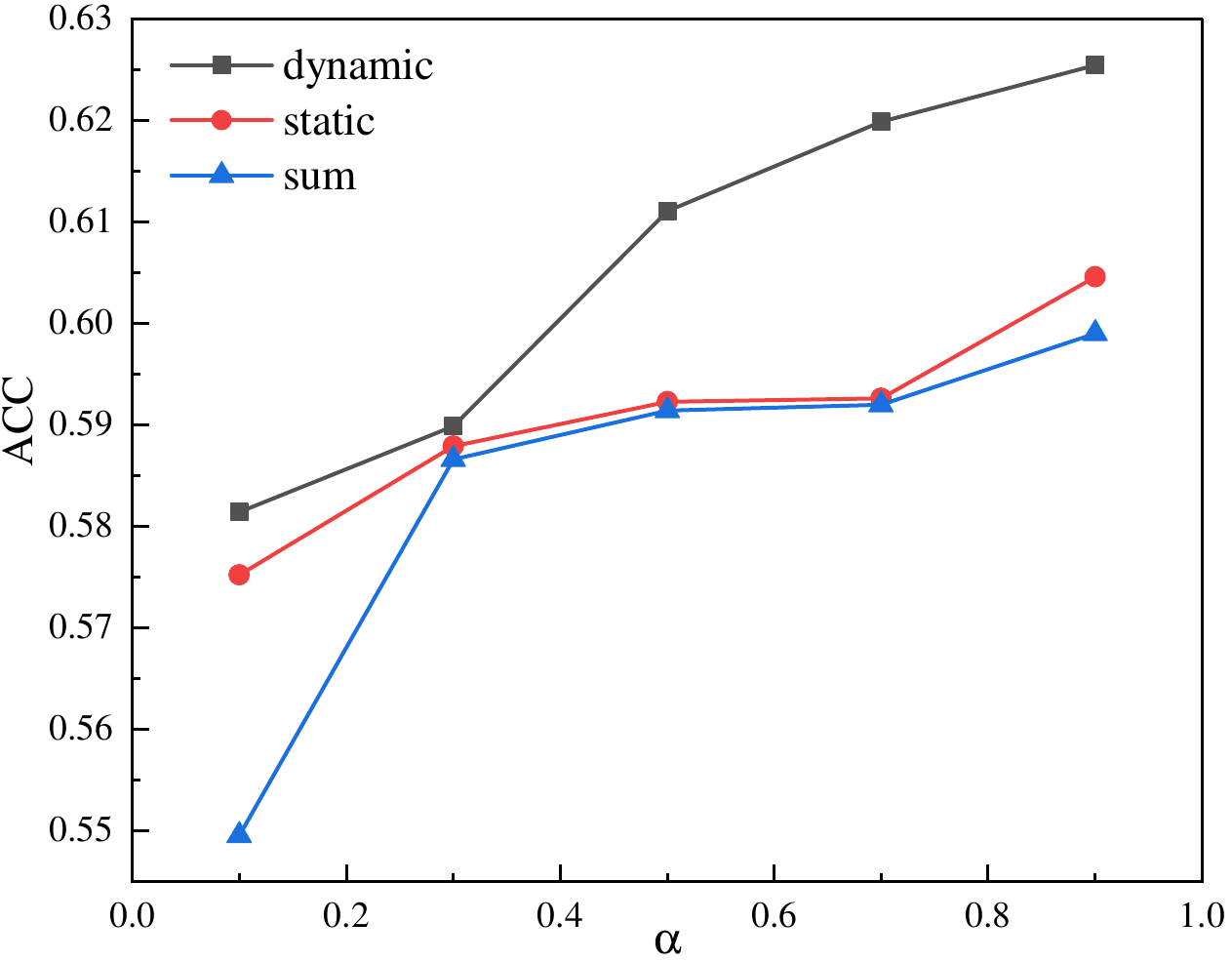}
			\caption{ LastFM-1k}
		\end{subfigure}
		\caption{Impact of the aggregation function on inference accuracy.}
		\label{agg}
	\end{figure}
\subsubsection{Aggregation methods in adaptive weight classification model}
 	 Items at different positions within the recommendation list may convey varying levels of users' information. To assess the impact of aggregating recommended item embedding on inference accuracy, a comparison is conducted with alternative aggregation methods, specifically direct summation and static weighted summation, which can be formulated as follows:
	\begin{align}
	    &Sum: u_i=\frac{1}{K} \times \sum_{j\in z_{i}  }^{}e_{j}^{S} \label{Eq: direct}\\ 
	    &Static: u_i=\frac{1}{K}\times \sum_{j\in z^{i} }^{} \frac{K-j+1}{K}\times  e_{j}^{S} \label{Eq: static}
	\end{align}
	The comparison results of the experiment are depicted in Figure \ref{agg}, which demonstrate the following:
	\begin{enumerate}
		\item The dynamic weighted summation method proves to be superior for aggregating the embedding of items within the recommendation list, as it yields higher attribute prediction accuracy compared to the other two aggregation methods across evaluation scenarios, verifying the hypothesis that items at various positions within the recommendation list have distinct impacts on the user attribute representations.
		\item The ultimate attribute prediction accuracy for each of the three aggregation methods increases with the rising proportion of participating users. This upward trend is due to the correlation between a larger percentage of participating users and a greater volume of supervised information for the classification model's training.
		\item The significant boost in attribute prediction accuracy as the proportion of participating users increases from 0.1 to 0.5 could be attributed to the enhanced supervision provided by the adaptive weight classification model for the user attributes of the newly included participating users, while the rest of the newly added attribute labels may include some noise.
	\end{enumerate} 
	
	\subsection{Analytical Capability of RAPI Framework against Recommendation Models Considering System Robustness (\textbf{RQ3})}
	\begin{table}[]
		\centering
		\caption{Analytical results against FairMI for gender across different scenarios in MovieLens-1m (Results within parentheses represent inference against LightGCN for gender in MovieLens-1m). }
		\begin{tabular}{c|c|cccccc}
			\hline
			\multirow{2}{*}{scenario} & \multirow{2}{*}{method} & \multicolumn{6}{c}{MovieLens-1m}                                                                 \\ \cline{3-8}
			&                         & \multicolumn{1}{c|}{$\alpha$}   & acc($\beta$=0.1) & acc($\beta$=0.3) & acc($\beta$=0.5) & acc($\beta$=0.7) & acc($\beta$=0.9) \\ \hline
			\multirow{3}{*}{1}        & DT                      & \multicolumn{1}{c|}{0}   & 0.5215 (0.5958)     & 0.5226 (0.5999)     & 0.5268 (0.6021)    & 0.5278 (0.6038)     &0.5283 (0.6041)     \\
			& KNN                     & \multicolumn{1}{c|}{0}   & 0.5317 (0.6170)    & 0.5324 (0.6175)    & 0.5339 (0.6248)    & 0.5347 (0.6447)     & 0.5349 (0.6497)     \\
			& DNN                     & \multicolumn{1}{c|}{0}   & 0.5664 (0.6801)     & 0.5692 (0.6976)     & 0.5711 (0.7014)     & 0.5723 (0.7086)     & 0.5729 (0.7103)     \\ \hline
			\multirow{3}{*}{2}        & DT                      & \multicolumn{1}{c|}{0}   & 0.5768 (0.6549)     & 0.5777 (0.6560)     & 0.5779 (0.6559)     & 0.5781 (0.6634)     & 0.5790 (0.6904)     \\
			& KNN                     & \multicolumn{1}{c|}{0}   & 0.6003 (0.6921)     & 0.6012 (0.6968)     & 0.6046 (0.6974)     & 0.6051 (0.6981)     & 0.6074 (0.6999)     \\
			& DNN                     & \multicolumn{1}{c|}{0}   & 0.6284 (0.7316)    &0.6311 (0.7336)     & 0.6357 (0.7342)    & 0.6366 (0.7355)     & 0.6379 (0.7378)     \\ \hline
			3                         & RAPI                     & \multicolumn{1}{c|}{0}   & 0.7168 (0.7345)     & 0.7177 (0.7358)   &0.7194 (0.7368)     & 0.7199 (0.7379)     & 0.7207 (0.7390)    \\ \hline
			\multirow{5}{*}{4}        & \multirow{5}{*}{RAPI}    & \multicolumn{1}{c|}{0.1} & 0.7324 (0.7517)     & 0.7348 (0.7529)     & 0.7369 (0.7556)     & 0.7383 (0.7557)     & 0.7392 (0.7585)    \\
			&                         & \multicolumn{1}{c|}{0.3} & 0.7346 (0.7539)     & 0.7348 (0.7546)     & 0.7356 (0.7550)   &0.7362 (0.7568)    & 0.7381 (0.7569)     \\
			&                         & \multicolumn{1}{c|}{0.5} & 0.7351 (0.7541)     & 0.7356 (0.7548)     & 0.7362 (0.7552)    & 0.7366 (0.7552)     & 0.7374 (0.7586)     \\
			&                         & \multicolumn{1}{c|}{0.7} & 0.7353 (0.7543)   & 0.7358 (0.7556) & 0.7364 (0.7558)     & 0.7366 (0.7569)    & 0.7368 (0.7570)     \\
			&                         & \multicolumn{1}{c|}{0.9} & 0.7364 (0.7614)     & 0.7367 (0.7623)      & 0.7372 (0.7631)     & 0.7379 (0.7634)    & 0.7385 (0.7640)     \\ \hline
		\end{tabular}
		\label{adversarial}
	\end{table}
	To systemically explore the potential impact of RAPI framework, RAPI is utilized to infer user attributes against recommendation model considering system robustness, e.g. FairMI. FairMI removes as much private information as possible and retains non-private information when generating embedding from a mutual information perspective to achieve the goal of fairness recommendation \cite{13}. Analytical results against FairMI are presented in table \ref{adversarial}, inference results against LightGCN is shown in parenthesess. Several observations can be summarized as follows:
	\begin{enumerate}
		\item In general, the analysis performance against FairMI lags behind that of LightGCN. This is due to the fact that FairMI is designed to produce fairness representations by minimizing the exposure of user behavioral patterns from a mutual information perspective, ensuring that the recommendation lists generated for users maintain balanced feature distribution to some extent.
		\item While FairMI offers some handling of user behaviors, its resilience to various analytical approaches is inconsistent. In scenarios 1 and 2, the performance of methods based on DT, KNN, and DNN typically deteriorates by about 8 percentage points, and can even experience a significant drop of up to 13 percentage points in more extreme cases. However, there is only a 2 to 3 percentage point drop in the performance of the RAPI method, indicating the effectiveness of the RAPI framework, even in the face of fairness-oriented recommendation models. This may be due to the fact that the RAPI framework is able to mine user attribute features by introducing content embedding of items obtained from pre-trained models as supplementary knowledge, which is independent of the original model, enabling more comprehensive user profile construction based on available user data and interaction information.

	\end{enumerate}

	\subsection{Performance of RAPI Framework Utilizing Advanced Pre-trained Models  (\textbf{RQ4})}
	
	\begin{figure}[]
		\centering
		\begin{subfigure}{0.49\linewidth}
			\includegraphics[width=\linewidth]{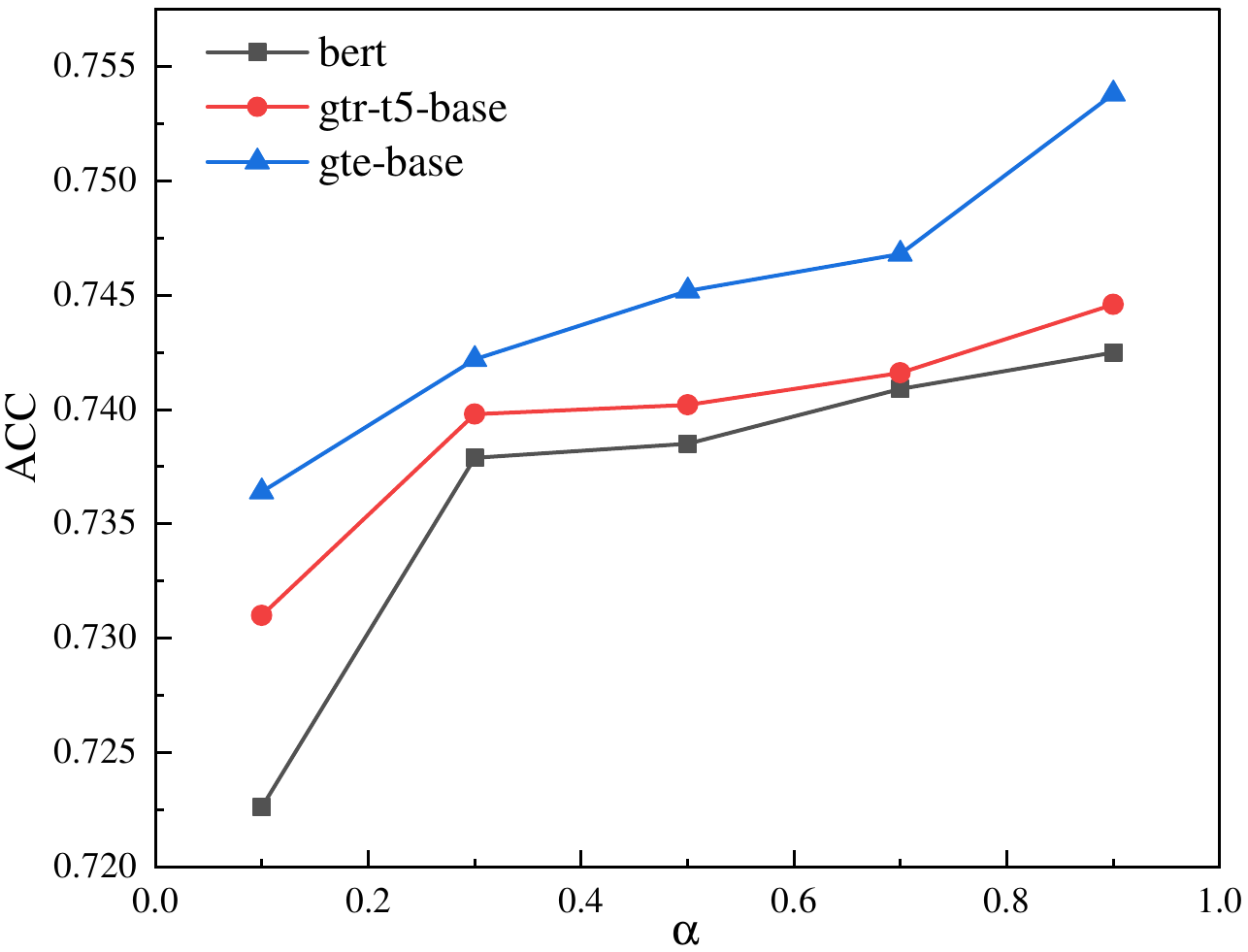}
			\caption{ MovieLens-1m}
		\end{subfigure}
		\hfill
		\begin{subfigure}{0.49\linewidth}
			\includegraphics[width=\linewidth]{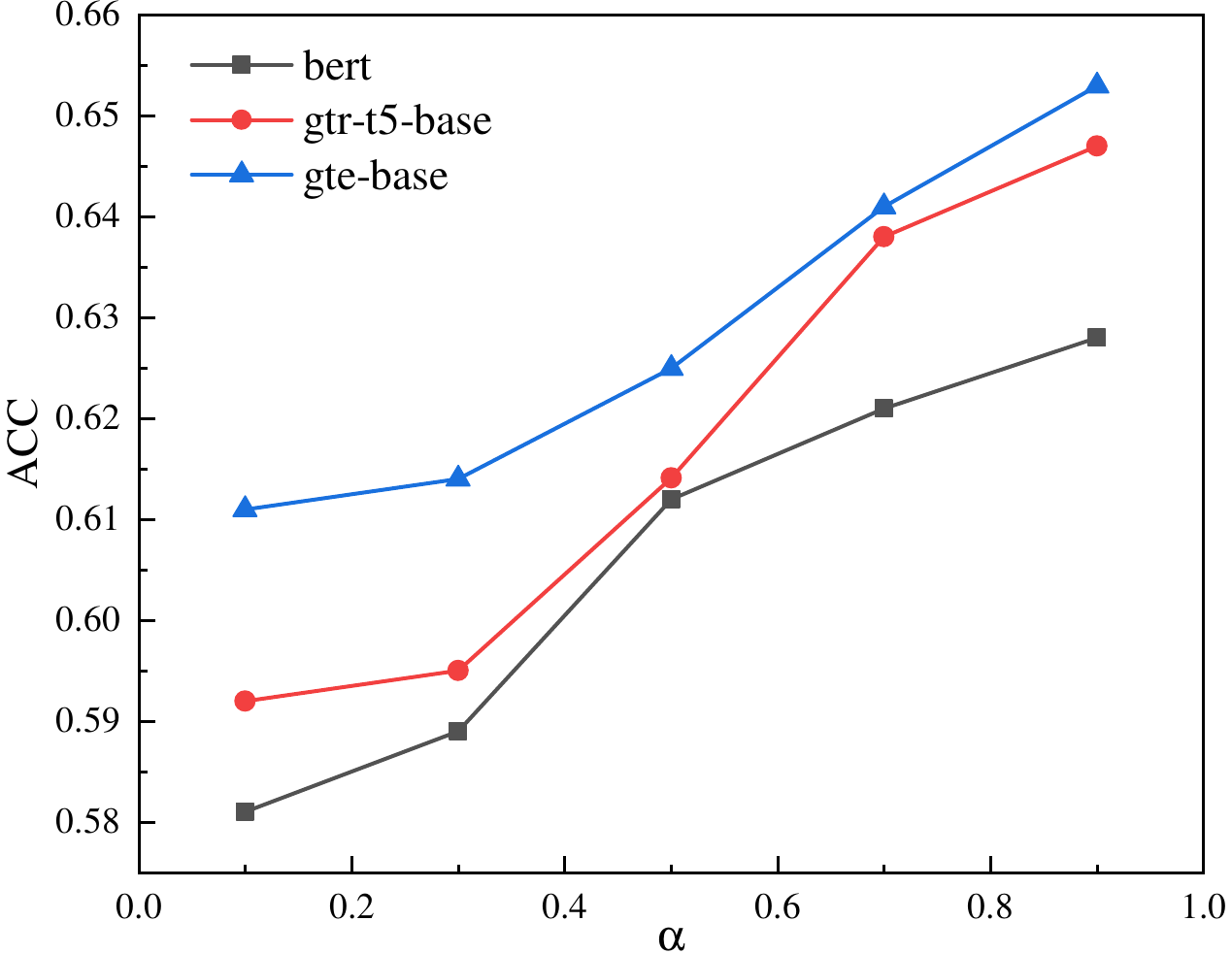}
			\caption{ LastFM-1k}
		\end{subfigure}
		\caption{Inference accuracy of RAPI framework utilizing various pre-trained models on MovieLens-1m and LastFM-1k.}
		\label{emb}
	\end{figure}
	
	In order to explore the effect of pre-trained model knowledge on attribute inference, the attribute inference results utilizing the advanced embedding model, rather than default BERT, are shown in table \ref{emb}. Advanced embedding model, gte-base and gtr-t5-base, are chosen to mine sensitive attribute embedding due to their powerful capabilities of representing words. gte-base is general text embedding model, which towards general text embeddings with multi-stage contrastive learning \cite{gte}. gtr-t5-base is a sentence-transformers model, which maps sentences to a 768 dimensional dense vector space \cite{gtr}. For a fair comparison, all content embedding extracted from the pre-trained model are 768 dimensions. Several observations can be summarized as follows:
	\begin{enumerate}
		\item The content embedding extracted from different embedding models have different impacts on the inference accuracy. The more suitable embedding model for extracting item content information will provide more supervisory information to RAPI framework. Specifically, gte-base can facilitate more precise inference of sensitive attributes within the RAPI framework compared to the gtr-t5-base and BERT. This enhanced accuracy may stem from the fact that gte-base benefits from a more comprehensive training dataset, while gtr-t5-base is better tailored for capturing holistic sentence features as opposed to the discrete item titles.
		\item The greater the proportion of users who provide interaction data, the more detailed the resulting user profiles become, regardless of the embedding model employed. An increase in available interaction data enables subsequent alignment models to learn more accurate recommendation embeddings, which in turn facilitates the construction of more precise and comprehensive user portraits based on behavioral patterns and item preferences.
		
	\end{enumerate}
	
	\subsection{Influence of the Length of the Augmented Recommendation List (\textbf{RQ5})}
	\begin{table}[]
		\centering
		\caption{Influence of the length of the augmented recommendation list.}
		\begin{tabular}{c|cccccccc}
			\hline
			\multirow{2}{*}{dataset} & \multicolumn{8}{c}{the length of the recommendation list} \\ \cline{2-9}
			& 50        & 60        & 70        & 80        & 90    & 100  &110 &120   \\ \hline
			MovieLens-1m                    & 0.7529    & 0.7534    & 0.7541    & 0.7543    & 0.7549  &0.7553 &0.7554 &0.7559 \\ \hline
			LastFM-1k                & 0.5723    & 0.5726    & 0.5734    & 0.5734    & 0.5738   &0.5741 &0.5743 & 0.5744\\ \hline
		\end{tabular}
		
		\label{reclength}
	\end{table}
	
	 This section is designed to assess the impact of the augmented recommendation list's length on the accuracy of user profile construction. In the experimental setup, the initial length of the recommendation list derived from user interactions is 20. The system can generate an augmented recommendation list of any length by performing list expansion techniques. The results of using augmented lists of different lengths for the final construction of user portraits are shown in table \ref{reclength}, and some findings can be summarized as follows:
	 \begin{enumerate}
	 	\item The precision of user profiling progressively enhances as the length of the augmented recommendation list expands. This upward trend is rational, considering that an extended list affords richer data for the profiling algorithm to discern patterns related to user preferences and behavioral characteristics, thus increasing the likelihood of achieving a more detailed and accurate user portrait.
	 	\item Furthermore, it is observable that an extension in the length of the original recommendation list exerts a minimal influence on the baseline capability to construct user profiles. This serves as an indication that even minor disclosures of recommendation lists can enable the system to generate user portraits with a certain level of precision. Concurrently, it becomes apparent that movie recommendations harbor more distinctive indicators of user preferences compared to music recommendations. This is evident when the accuracy of user profile construction for the MovieLens-1m dataset outperforms that of Last-FM by a substantial 18 percentage points, under identical experimental conditions, suggesting that movie interaction data provides stronger signals for (depicting) comprehensive user characteristics.
	 \end{enumerate}
	
	\subsection{Complexity Analysis}

After validating the efficacy of the RAPI framework, a comprehensive complexity analysis is conducted.
Especially, the time and space complexity of RAPI framework are as follows:
	\begin{align}
	&Time \; complexity: \mathcal{O}  \left ( K*(L_1*L_2+L_3)\right )  +\mathcal{O}  \left ( L_3*K*S  \right ) \\ 
	&Space \; complexity:\mathcal{O}  \left ( (L_1+L_2+L_3)*(M+N)  \right ) ) 
\end{align}
where $L_1$, $L_2$ and $L_3$ represent the size of item embedding obtained from 
origin recommendation model, pre-trained language model and spurious model, respectively. $K$ represent the length of original recommendation lists obtained by analyst. $M$ and $N$ represent number of users and items, respectively.
The first term of its time complexity primarily stems from the first two steps of the RAPI framework, spurious model confirmation and data augmentation for recommendation lists, which are designed to augment the recommendation lists for  users and improve inference attributes capability of RAPI.

\subsection{Strategy for Enhancing User Profile Robustness}\label{defend}
	While the experiments confirm the effectiveness of our RAPI framework in constructing user profiles, developing a robust mechanism to ensure profile stability under data variations remains an open challenge.
	Inspired by the insight that introducing variations in recommendation lists can affect the precision of profile construction, partial items in recommendation lists are replaced with others of the same category; experimental results are shown in table \ref{defense}. It is observed that the accuracy of the RAPI framework in depicting user characteristics exhibits varying degrees of fluctuation under this strategy. However, there is no doubt that introducing variations in recommendation lists would lead to a decline in overall recommendation performance. Future research will focus on optimally adjusting recommendation lists to maximize the robustness of user profile construction against data noise while preserving high recommendation accuracy.
	
    \begin{table}[!h]
		\caption{Inference of RAPI under Robustness Strategy (Results without strategy in parentheses).}
		\begin{tabular}{cccc}
			\toprule
			Datasets      & Gender         & Age            & Occupation     \\ \midrule
			MovieLens-1m  & 0.7486 (0.7517) & 0.3691 (0.3714) & 0.1511 (0.1543) \\
			LastFM-1k     & 0.5692 (0.5723) & --              & --              \\
			Book-Crossing & --              & 0.2311 (0.2346) & --              \\
			Cold-Rec      & 0.6712 (0.6725) & --              & --              \\ \bottomrule
			\label{defense}
		\end{tabular}
	\end{table}
	
	\section{Conclusion and Future Work}
	In this paper, we establish various information utilization scenarios and evaluate the potential for constructing detailed user profiles from recommendation lists. We introduce a comprehensive framework, RAPI, which consists of three core components: first, a surrogate recommendation model is developed based on list similarity to simulate the behavior of the original system; second, an augmentation module leverages recommendation embedding similarity to retrieve comprehensive item representations, thereby extending the target user's recommendation list; third, an adaptive weight classification model dynamically assigns weights to item embeddings based on their positions within the list. Extensive experiments on real-world benchmarks demonstrate that RAPI achieves high precision in user profiling using available interaction records.

    To enhance the robustness of our system, future work will investigate optimal strategies to maintain profile stability against input perturbations and noise. Specifically, we will explore refinement mechanisms for recommendation lists to achieve an optimal trade-off between profiling precision and recommendation performance. This includes determining the appropriate variation intensity and application scope to mitigate informational inconsistencies while minimizing potential degradation in recommendation quality.
		
	\bibliography{fu}
 \end{document}